\documentclass[12pt,a4paper,aps,prd,preprint,superscriptaddress,nofootinbib,preprintnumbers]{revtex4-1}
\usepackage[utf8]{inputenc}
\usepackage{graphicx}
\usepackage{amssymb}
\usepackage{textcomp}
\usepackage{amsmath}
\usepackage{tabularx}
\usepackage{bm}
\usepackage{times}
\usepackage{color}
\usepackage{slashed}
\usepackage{multirow}
\usepackage{verbatim}
\usepackage{ulem}

\usepackage[colorlinks=true, pdfstartview=FitV, linkcolor=blue, citecolor=blue, urlcolor=blue]{hyperref}
\allowdisplaybreaks[4]

\def\lsim{\mathrel{\raise.3ex\hbox{$<$\kern-.75em\lower1ex\hbox{$\sim$}}}}
\def\gsim{\mathrel{\raise.3ex\hbox{$>$\kern-.75em\lower1ex\hbox{$\sim$}}}}

\newcommand{\calO}{{\mathcal{O}}}

\newcommand{\Tr}{{\rm Tr}}

\definecolor{orange}{rgb}{1,0.5,0}

\begin{document}

\title{The implication of $J/\psi\to (\gamma + ){\rm invisible}$ for the effective field theories of neutrino and dark matter}

\author{Tong Li}
\email{litong@nankai.edu.cn}
\affiliation{
School of Physics, Nankai University, Tianjin 300071, China
}
\author{Xiao-Dong Ma}
\email{maxid@sjtu.edu.cn}
\affiliation{
Tsung-Dao Lee Institute, and School of Physics and Astronomy, Shanghai Jiao Tong University, Shanghai 20024, China}

\author{Michael A. Schmidt}
\email{m.schmidt@unsw.edu.au}
\affiliation{
Sydney Consortium for Particle Physics and Cosmology,
School of Physics, The University of New South Wales, Sydney, New South Wales 2052, Australia
}
\author{Rui-Jia Zhang}
\email{zhangreka1998@sina.com}
\affiliation{
School of Physics, Nankai University, Tianjin 300071, China
}

\preprint{CPPC-2021-05}
\begin{abstract}
We study the implication of $J/\psi$ decay into invisible particles for light sterile neutrino and sub-GeV dark matter (DM).
The low-energy effective field theories (EFTs) are used for the description of general neutrino interactions and the Dirac fermion DM coupled to charm quark.
For $J/\psi\to \gamma+{\rm invisible}$, we perform the likelihood fits for the individual neutrino and DM operators with distinct Lorentz structures and photon spectra.
The limits on the decay branching fractions are obtained for different neutrino or DM scenarios and then converted to the lower bounds on the new energy scales.
The most stringent bounds on the energy scale in neutrino and DM EFTs are 12.8 GeV and 11.6 GeV, respectively. The purely invisible decay $J/\psi\to {\rm invisible}$ provides complementary constraints on the effective operators. The relevant bound on the energy scale is above 100 GeV for the dipole operators.
We also evaluate the limit on the DM-nucleon scattering cross section converted from $J/\psi$ data. The data of $J/\psi$ invisible decays are sensitive to the light DM mass range where the DM direct detection experiments cannot probe yet. The future Super Tau Charm
Factory after one year run can push the limits down by two orders of magnitude.
\end{abstract}

\maketitle

\section{Introduction}
\label{sec:Intro}

The observation of neutrino flavor oscillations~\cite{Fukuda:1998mi}
requires non-vanishing neutrino masses and thus
provides a strong motivation for new physics (NP) beyond the Standard Model (SM) associated with neutrinos.
Meanwhile, abundant cosmological and astrophysical observations clearly hint towards
the existence of dark matter (DM) as a plausible new physics beyond the SM.
However, no convincing signal has been observed yet for any electroweak-scale DM candidates and neither for new dynamical degrees of freedom in the neutrino sector such as right-handed (RH) neutrinos $N$.
This motivates to broaden the scope and thus
much attention is recently being paid to probe light DM and light sterile neutrinos.
Some anomalies in direct DM detection or short-baseline neutrino oscillation hint the existence of
light DM~\cite{Aprile:2020tmw,Zhou:2020bvf} or light sterile neutrino states (see Refs.~\cite{Diaz:2019fwt,Boser:2019rta} and references therein).
The detection of DM scattering off electrons or the precision measurements in neutrino experiments may reveal possible NP associated with light degrees of freedom in the SM.

Besides the experiments sensitive to electrons or light SM quarks, the heavy quarkonium experiments provide an ideal environment to study the possible NP associated with heavy quarks. The CLEO-c~\cite{Insler:2010jw}, BaBar~\cite{delAmoSanchez:2010ac} and Belle~\cite{Seong:2018gut} experiments have searched for $J/\psi$ or $\Upsilon$ radiative decays into invisible particles.
Recently, the BESIII collaboration performed a similar search for $J/\psi$ using the data collected by the BESIII detector at Beijing Electron Positron Collider (BEPCII), and no signal was observed~\cite{Ablikim:2020qtn}. The BESIII collaboration interprets the invisible particle as a new CP-odd pseudoscalar and the upper limit on the branching fraction for a massless pseudoscalar is $7\times 10^{-7}$ at the 90\% confidence level (C.L.). The experimental sensitivity does not reach
the SM prediction of $\mathcal{B}(J/\psi\to \gamma\nu\bar{\nu})\simeq 7\times 10^{-11}$~\cite{Gao:2014yga,Bai:2017wle}, but the BESIII data is
sensitive to the low-energy NP interacting with SM charm quark.

The $J/\psi$ decays into a single photon together with invisible particles is analogous to the mono-photon signature at high-energy colliders and can also be used to search for light sterile neutrino or sub-GeV DM.
As usually protected by a global symmetry, the DM particles are generally produced in pairs. The neutral current interactions of neutrinos also lead to their pair productions. We thus investigate the three-body decays of $J/\psi$ into a photon and a pair of neutrinos or DM particles, rather than the two-body decay of $J/\psi$ which has been analyzed by BESIII~\cite{Ablikim:2020qtn}.
Similar studies inspired by the Belle data were performed for the $\Upsilon$ radiative decay into DM pairs~\cite{Yeghiyan:2009xc,Fernandez:2015klv} and the invisible decay of dark photon or millicharged particle was also proposed through the $e^+e^-$ collision at the BESIII detector~\cite{Liu:2018jdi,Zhang:2019wnz}. There were also a number of theoretical studies of the search for light DM using invisible quarkonium decays~\cite{Fayet:2007ua,McElrath:2007sa,Fayet:2009tv,McKeen:2009rm,Essig:2013vha,Cotta:2013jna,Fernandez:2014eja,Bertuzzo:2017lwt,Bertuzzo:2020rzo}
at $e^+e^-$ colliders.
Another benefit of the heavy quarkonium decays to a photon and invisible particles is the ability to probe arbitrarily small DM masses. This is in contrast to low-energy DM-nucleon scattering experiments which lose sensitivity when the recoil energy becomes too small to reach the energy threshold of DM direct detection experiments.
Hence, the study of heavy quarkonium decays allows to probe parameter space which is inaccessible to DM direct detection experiments.

For the low-energy processes involving neutrinos and DM particles, the effective field theory (EFT) serves as a model-independent framework to study the implications
for neutrino physics and DM without recourse to detailed NP models. The low-energy effective field theory (LEFT) is an EFT for the light SM quarks and leptons defined
below the electroweak scale and is valid above the chiral symmetry breaking scale $\simeq 1$ GeV for the interactions involving quarks. The LEFT respects SU$(3)_{\rm c}\times$U$(1)_{\rm em}$ gauge symmetry and can well describe the low-energy physics in the heavy quarkonium experiments. The LEFT Lagrangian is
\begin{align}
\mathcal{L}_{\rm LEFT}=\mathcal{L}_{\rm d\leq 4}+ \sum_i \sum_{d\geq 5} C_{i}^{(d)} \calO_{i}^{(d)}\;,
\end{align}
where $C_{i}^{(d)}$ is the Wilson coefficient (WC) of operator $\calO_{i}^{(d)}$. We make use of the LEFT with RH neutrinos $N$ named as LNEFT~\cite{Chala:2020vqp,Li:2020lba,Li:2020wxi} and the LEFT with DM~\cite{Beltran:2008xg,Fan:2010gt,Goodman:2010qn,Balazs:2014rsa,DeSimone:2016fbz,Brod:2017bsw,Fitzpatrick:2012ix,Fitzpatrick:2012ib,DelNobile:2013sia,Bishara:2017pfq,Bischer:2020sop} called DMEFT below. Generally, the Wilson coefficient $C_{i}^{(d)}$ scales as $\Lambda^{4-d}$.
The constraint on the effective energy scale $\Lambda$ is generally related to the mediator mass $m_{\rm med}\sim g\Lambda$ in UV completions with $g$ being a coupling for a given interaction. The EFT approximation is valid when the mediator mass is sufficiently larger than the momentum transfer in the experimental processes. Assuming $g\sim \mathcal{O}(1)$, the mono-photon searches at high-energy colliders can only set bounds for the new energy scale above TeV scale for a valid EFT description. Here, from heavy quarkonium $J/\psi$ decay, the validity of the EFT description is ensured for the limits larger than about 3 GeV. If $g$ is decreased to a smaller value, the momentum transfer $\sqrt{\hat{s}}$ would dominate over the mediator mass for $m_{\rm med}\ll \sqrt{\hat{s}}$ and the mono-photon event rate is suppressed by $g^4/\hat{s}^2$. By contrast, the event rate from heavy quarkonium decay is proportional to $g^4/m_{\rm med}^4$ and the constraint would stay constant. Thus, in the EFT frameworks we consider, the heavy quarkonium decays can provide complementary constraints on the NP scale in which the searches at high-energy colliders are not applicable or lose sensitivity. Following the likelihood fit performed on the photon energy range from 1.25 to 1.65 GeV by BESIII, we perform the fits for the individual LNEFT and DMEFT operators with distinct Lorentz structures and photon spectra. The limits on the new scale $\Lambda$ can be obtained and then converted into the bound on DM-nucleon scattering cross section in DMEFT.

The physics potential of current BEPCII/BESIII is limited by its
luminosity and the center-of-mass energy. A Super Tau Charm Facility (STCF) is proposed as a natural extension and a viable option for an accelerator based
high energy project in China in the post BEPCII/BESIII era~\cite{STCF0}. It is designed to have
c.m. energy ranging from 2 to 7 GeV, and is expected to deliver more than 1 ab$^{-1}$ of integrated luminosity per year. For comparison, the Belle II experiment is expected to accumulate 50 ab$^{-1}$ data by 2024~\cite{Belle-II:2018jsg} and the LHCb will also collect much more data in future~\cite{LHCb:2018roe}. Although the STCF might be at a disadvantage in terms of the absolute number of events, it has an excellent signal to background ratio, high detection efficiency, well-controlled systematic uncertainties, capabilities for fully
reconstructed event, and it provides an excellent opportunity for a broad range of physics studies in the tau-charm energy region.
We also provide the sensitivity on DM-nucleon scattering at future STCF.
The results from heavy quarkonium experiments can guide our direct search for NP in the neutrino or DM sector in future experiments.

Besides the radiative decay, the purely invisible decay of heavy quarkonium can also place constraints on the effective operators in LNEFT and DMEFT~\cite{Fayet:1979qi,Fayet:1991ux,McElrath:2005bp,Chang:1997tq}. The first search for $J/\psi$ decay to invisible final states gave the 90\% C.L. upper limit $\mathcal{B}(J/\psi\to {\rm invisible})<1.2\times 10^{-2}\mathcal{B}(J/\psi\to \mu^+\mu^-)$~\cite{Ablikim:2007ek} which is quoted as $\mathcal{B}(J/\psi\to {\rm invisible})<7\times 10^{-4}$ in PDG~\cite{Zyla:2020zbs}.
As the invisible decay has no suppressions from the QED vertex and the 3-body phase space in the radiative decay, one expects that it would bring more stringent bounds on the new energy scale. On the other hand, due to the C parity conservation in the decay of $J/\psi$ with $J^{C}=1^{-}$, $J/\psi\to \gamma+{\rm invisible}$ and $J/\psi\to {\rm invisible}$ are respectively induced by C-even and C-odd operators. Thus, they provide complementary constraints to the heavy quarkonium interactions with RH neutrinos or DM particles.

The paper is outlined as follows. In Sec.~\ref{sec:EFT}, we describe the EFT frameworks for general neutrino interactions with RH neutrinos and Dirac fermion DM. We then calculate the heavy quarkonium radiative decay into invisible particles in Sec.~\ref{sec:decay} and the purely invisible decays in Sec.~\ref{sec:invdecay}. In Sec.~\ref{sec:Num}, we show the numerical constraints on the decay branching fractions and the NP scale in both LNEFT and DMEFT. Our conclusions are summarized in Sec.~\ref{sec:Con}. The details of our calculation are presented in the Appendices.

\section{Effective Field Theories of neutrino and dark matter}
\label{sec:EFT}

\subsection{General neutrino interactions with RH neutrinos}
\label{sec:GNI}

For the radiative decay of $J/\psi\to \gamma+\rm invisible$, it can be suitably investigated in the LNEFT framework in which the invisible states are neutrinos. The LNEFT is a valid description for physical processes taking
place below the electroweak scale $\Lambda_{\text{EW}}=m_W$. Its dynamical degrees of freedom include the SM light leptons $(e,\mu,\tau,\nu_e,\nu_\mu,\nu_\tau)$ and quarks $(u,d,s,c,b)$ and an arbitrary number of RH neutrinos $N$. The LNEFT Lagrangian consists of the
higher dimensional operators
built out of those fields and satisfies the gauge symmetry SU$(3)_{\rm c}\times $U$(1)_{\rm em}$.
The complete and independent operator basis involving RH neutrinos $N$ up to dim-6 in the LNEFT can be found in Refs.~\cite{Chala:2020vqp,Li:2020lba,Li:2020wxi} for the study of generic neutrino interactions.

The leading order LNEFT operators for the study of $J/\psi\to\gamma+\rm invisible$ decay are at dim-5 and dim-6~\footnote{Since the Wilson coefficients of the dim-7 operators~\cite{Liao:2020zyx} are usually suppressed by one more power of heavy scale than those of the dim-6 operators, we thus neglect the dim-7 LNEFT operators for the current work.}. They are composed of a neutrino bilinear coupled to the photon field strength tensor (for the dim-5 case) or SM quark bilinear currents (for the dim-6 case). Those operators are further classified in terms of whether or not the lepton number is violated. For the lepton number conservation (LNC, $|\Delta L|=0$) case, the dim-5 neutrino-photon and dim-6 neutrino-quark operators are given by~\cite{Jenkins:2017jig,Li:2020lba}
\begin{align}
\calO_{\nu N F}&=(\overline{\nu}\sigma_{\mu\nu}N) F^{\mu\nu} +h.c.\; ,
\label{LNCdipole}
\\
\calO_{q\nu 1}^{V}&=(\overline{q_L}\gamma_\mu q_L)(\overline{\nu}\gamma^\mu\nu)\; , &
\calO_{q\nu 2}^{V}&=(\overline{q_R}\gamma_\mu q_R)(\overline{\nu}\gamma^\mu\nu)\; ,
\label{LNCqnu}
\\
\calO_{qN 1}^{V}&=(\overline{q_L}\gamma_\mu q_L)(\overline{N}\gamma^\mu N)\; , &
\calO_{qN 2}^{V}&=(\overline{q_R}\gamma_\mu q_R)(\overline{N}\gamma^\mu N)\; ,
\label{LNCqN}
\\
\calO_{q\nu N1}^S&= (\overline{q_L} q_R)(\overline{\nu} N)+h.c.\; ,&
\calO_{q\nu N2}^S&= (\overline{q_R} q_L)(\overline{\nu} N)+h.c.\; ,
\label{LNCqnuN}
\\
\calO_{q\nu N}^T&= (\overline{q_L} \sigma^{\mu\nu}q_R)(\overline{\nu}\sigma_{\mu\nu} N)+h.c.\; ,
\label{LNCqnuNT}
\end{align}
where $F_{\mu\nu}$ is the electromagnetic field strength tensor, $q$ can be either up-type quarks $u_i=(u,c)$ or down-type quarks $d_i=(d,s,b)$, $\nu_i$ are active left-handed neutrinos $(\nu_e,\nu_\mu, \nu_\tau)$, and $N_i$ are RH neutrinos. The quark fields and the RH neutrino fields are in the mass basis, while the LH neutrino fields are in the flavor basis.
Both $\nu_i$ and $N_i$ carry lepton number $L(\nu_i)=L(N_i)=+1$.
The flavors of the two quarks and those of the two neutrinos in the above operators can be different although we do not specify their flavor indices here.
For the notation of the Wilson coefficients, we use the same subscripts as the operators, for instance $C_{q\nu1}^{V,pr\alpha\beta}$ together with $\calO_{q\nu1}^{V,pr\alpha\beta}$, where $p,r$ denote the quark flavors and $\alpha,\beta$ are the neutrino flavors.
We do not include `$h.c.$' for the vector-like operators in Eqs.~(\ref{LNCqnu},\ref{LNCqN}) because they are self-hermitian after exhausting all flavor indices.

The relevant dim-5 and dim-6 operators which induce lepton number violation (LNV, $|\Delta L|=2$) are
\begin{align}
\calO_{\nu \nu F}&=(\overline{\nu^C}\sigma_{\mu\nu}\nu) F^{\mu\nu} +h.c.\; , &
\calO_{NN F}&=(\overline{N^C}\sigma_{\mu\nu}N) F^{\mu\nu} +h.c.\; ,
\\
\calO_{q\nu N1}^V&= (\overline{q_L}\gamma_\mu q_L)(\overline{\nu^C}\gamma^\mu N)+h.c.\; , &
\calO_{q\nu N2}^V&=(\overline{q_R}\gamma_\mu q_R)(\overline{\nu^C}\gamma^\mu N)+h.c.\; ,
\label{LNVqnuN}
\\
\calO_{q\nu 1}^S&= (\overline{q_R} q_L)(\overline{\nu^C} \nu)+h.c.\; ,&
\calO_{q\nu 2}^S&= (\overline{q_L} q_R)(\overline{\nu^C} \nu)+h.c.\; ,
\label{LNVqnu}
\\
\calO_{qN 1}^S&= (\overline{q_R} q_L)(\overline{N^C} N)+h.c.\; ,&
\calO_{qN 2}^S&= (\overline{q_L} q_R)(\overline{N^C} N)+h.c.\; ,
\label{LNVqN}
\\
\calO_{q\nu}^T&= (\overline{q_R} \sigma^{\mu\nu}q_L)(\overline{\nu^C}\sigma_{\mu\nu} \nu)+h.c.\;, &
\calO_{qN}^T&= (\overline{q_L} \sigma^{\mu\nu}q_R)(\overline{N^C}\sigma_{\mu\nu} N)+h.c.\; .
\label{LNVT}
\end{align}
Note that the Wilson coefficients of the scalar operators are symmetric in the neutrino indices and the dipole and tensor operators are antisymmetric.
Thus in particular the operators with the tensor neutrino current $\overline{\nu_\alpha^C} \sigma^{\mu\nu} \nu_\beta$ or $\overline{N_\alpha^C} \sigma^{\mu\nu} N_\beta$ vanish for identical neutrino flavors ($\alpha=\beta$).

\subsection{Low-energy description of fermionic DM}
\label{sec:DMEFT}

The low-energy interactions of Dirac fermion DM $\chi$ we consider are based on the basis composed of the effective operators of DM and SM particles up to dim-7 given in~\cite{Brod:2017bsw}. A conserved global U$(1)$ symmetry is assumed in the dark sector to stabilize the DM. The two dim-5 dipole operators are
\begin{align}
\mathcal{O}_{\chi F1}&=(\overline{\chi}\sigma_{\mu\nu}\chi)F^{\mu\nu}\;,
&\mathcal{O}_{\chi F2}&=(\overline{\chi}\sigma_{\mu\nu}i\gamma_5\chi)F^{\mu\nu}\;.
\label{eq: dmF}
\end{align}
The four-fermion interactions between $\chi$ and SM quarks consist of dim-6 operators
\begin{align}
\mathcal{O}_{\chi q1}&=(\overline{\chi}\gamma_\mu\chi)(\overline{q}\gamma^\mu q)\;,
&\mathcal{O}_{\chi q2}&=(\overline{\chi}\gamma_\mu \gamma_5\chi)(\overline{q}\gamma^\mu q)\;,\\
\mathcal{O}_{\chi q3}&=(\overline{\chi}\gamma_\mu\chi)(\overline{q}\gamma^\mu \gamma_5 q)\;,
&\mathcal{O}_{\chi q4}&=(\overline{\chi}\gamma_\mu \gamma_5\chi)(\overline{q}\gamma^\mu \gamma_5 q)\;,
\end{align}
as well as the dim-7 operators with scalar or tensor currents
\begin{align}
\mathcal{O}_{\chi q5}&=m_q(\overline{\chi} \chi)(\overline{q}q)\;,
& \mathcal{O}_{\chi q6}&=m_q(\overline{\chi}i\gamma_5\chi)(\overline{q} q)\;,\\
\mathcal{O}_{\chi q7}&=m_q(\overline{\chi}\chi)(\overline{q}i\gamma_5 q)\;,
& \mathcal{O}_{\chi q8}&=m_q(\overline{\chi}i\gamma_5\chi)(\overline{q}i\gamma_5 q)\;,\\
\mathcal{O}_{\chi q9}&=m_q(\overline{\chi}\sigma_{\mu\nu}\chi)(\overline{q}\sigma^{\mu\nu} q)\;,
&\mathcal{O}_{\chi q10}&=m_q(\overline{\chi}\sigma_{\mu\nu}i\gamma_5\chi)(\overline{q}\sigma^{\mu\nu} q)\;,
\end{align}
and those with derivative in the DM current
\begin{align}
\mathcal{O}_{\chi q11}&=(\overline{\chi}i\overleftrightarrow{\partial_\mu} \chi)(\overline{q}\gamma^\mu q)\;, &\mathcal{O}_{\chi q12}&=(\overline{\chi}i\gamma_5 i\overleftrightarrow{\partial_\mu}\chi)(\overline{q}\gamma^\mu q)\;,\\
\mathcal{O}_{\chi q13}&=(\overline{\chi}i\overleftrightarrow{\partial_\mu} \chi)(\overline{q}\gamma^\mu \gamma_5 q)\;,
&\mathcal{O}_{\chi q14}&=(\overline{\chi}i\gamma_5 i\overleftrightarrow{\partial_\mu}\chi)(\overline{q}\gamma^\mu \gamma_5 q)\;,\\
{\cal O}_{\chi q15}&=(\overline{\chi}\gamma^{[\mu}i\overleftrightarrow{\partial}^{\nu]}\chi)(\overline{q}\sigma_{\mu\nu} q)\;,
&{\cal O}_{\chi q16}&=(\overline{\chi}\gamma^{[\mu}i\overleftrightarrow{\partial}^{\nu]}\gamma_5\chi)(\overline{q}\sigma_{\mu\nu} q)\;,
\\
{\cal O}_{\chi q17}&=(\overline{\chi}\gamma^{[\mu}i\overleftrightarrow{\partial}^{\nu]} \chi)(\overline{q}\sigma_{\mu\nu}i \gamma_5 q)\;,
&{\cal O}_{\chi q18}&=(\overline{\chi}\gamma^{[\mu}i\overleftrightarrow{\partial}^{\nu]}\gamma_5\chi)(\overline{q}\sigma_{\mu\nu}i\gamma_5 q)\;,
\label{eq: dmq}
\end{align}
where $\overline{\chi}i\overleftrightarrow{\partial_\mu} \chi=\overline{\chi}i\partial_\mu \chi - \overline{\chi}i \overleftarrow{\partial_\mu}\chi$.
We replaced the last four operators in Ref.~\cite{Brod:2017bsw}~\footnote{We slightly change the notation of the four operators. Their original notation in Ref.~\cite{Brod:2017bsw} is ${\cal O}_{19,q}^{(7)}$, ${\cal O}_{20,q}^{(7)}$, ${\cal O}_{21,q}^{(7)}$ and ${\cal O}_{22,q}^{(7)}$ in which the quark field label $q$ can be other SM lepton fields $\in\{e,\mu,\tau,\nu_e,\nu_\mu,\nu_\tau\}$.}
\begin{align}\nonumber
\hat{\mathcal{O}}_{\chi q15}&=\partial_\mu(\overline{\chi}\sigma^{\mu\nu}\chi)(\overline{q}\gamma_\nu q)\;, &
\hat{\mathcal{O}}_{\chi q16}&=\partial_\mu(\overline{\chi}\sigma^{\mu\nu}i\gamma_5\chi)(\overline{q}\gamma_\nu q)\;,\\
\hat{\mathcal{O}}_{\chi q17}&=\partial_\mu(\overline{\chi}\sigma^{\mu\nu}\chi)(\overline{q}\gamma_\nu \gamma_5 q)\;, &
\hat{\mathcal{O}}_{\chi q18}&=\partial_\mu(\overline{\chi}\sigma^{\mu\nu}i\gamma_5\chi)(\overline{q}\gamma_\nu \gamma_5 q)\;,
\label{eq: red_opes}
\end{align}
because they are redundant, by four independent operators $\mathcal{O}_{\chi q15-18}$.
The operators $\hat{\mathcal{O}}_{\chi q15-18}$ listed in Ref.~\cite{Brod:2017bsw} can be reduced into the operators within $\mathcal{O}_{\chi q1-14}$ using the Dirac gamma matrix identity (GI) $\sigma^{\mu\nu}={i\over 2}[\gamma^\mu,\gamma^\nu]=i\gamma^\mu\gamma^\nu-ig^{\mu\nu}=ig^{\mu\nu}-i\gamma^\nu\gamma^\mu$ and the equation of motion (EoM) of the DM fields.
We prove the redundancy of operators $\hat{\cal O}_{\chi q15,16,17,18}$ and the above realization in Appendix~\ref{sec:DMredundant}.

The dim-7 Rayleigh operators with the DM coupled to two photon field strength tensors may also contribute to the radiative decay of interest
\begin{align}
\mathcal{O}_{\chi FF1}&=(\overline{\chi} \chi) F^{\mu\nu} F_{\mu\nu}\;,
&\mathcal{O}_{\chi FF2}&=(\overline{\chi}i\gamma_5 \chi) F^{\mu\nu} F_{\mu\nu}\;,\\
\mathcal{O}_{\chi FF3}&=(\overline{\chi} \chi) F^{\mu\nu} \tilde{F}_{\mu\nu}\;,
&\mathcal{O}_{\chi FF4}&=(\overline{\chi}i\gamma_5 \chi) F^{\mu\nu} \tilde{F}_{\mu\nu}\;.
\end{align}
Note that the dim-7 operators with gluon field strength tensors are irrelevant for our study and are not listed here.

\section{The heavy quarkonium radiative decay into invisible particles}
\label{sec:decay}

\subsection{The radiative decay of heavy quarkonium in LNEFT}
\label{sec:decayinLNEFT}

\begin{figure}[h!]
\centering
\includegraphics[width=15cm]{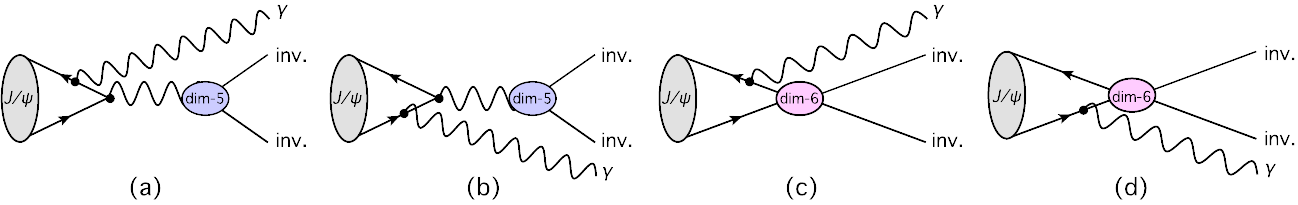}
\caption{The Feynman diagrams contributing to $J/\psi\to \gamma+{\rm inv}_1+{\rm inv}_2$ (${\rm inv}_{1,2}\in\{\nu,\bar\nu,N,\bar N \}$) process from the dim-5 (a,b) and dim-6 (c,d) EFT interactions in the LNEFT. }
\label{Fig1}
\end{figure}

For the $J/\psi$ radiative decays into invisible fermions, the leading order Feynman diagrams are shown in Fig.~\ref{Fig1} in the framework of the LNEFT. The relevant local operators are defined in the above section and the photon is only emitted from the quark lines. The transition amplitude for $J/\psi(P)\to \gamma(k)+{\rm inv}_1(k_1)+{\rm inv}_2(k_2)$ through the diagrams in Fig.~\ref{Fig1} can be factorized into a hadronic matrix element ${\cal  H}^{\Gamma}$ multiplied by a proper neutrino current $L_{\rm inv}$
\begin{align}
{\cal M}(J/\psi\to\gamma + {\rm inv}_1+{\rm inv}_2)=\langle\gamma(k)|(\bar q\Gamma q)|J/\psi(P)\rangle\times L_{\rm inv}\equiv{\cal  H}^{\Gamma}L_{\rm inv}\;,
\end{align}
where $(\bar q\Gamma q)$ represents a generic quark current appearing in the local dim-6/7 interactions with
$\Gamma=\{P_\pm\equiv {1\pm\gamma_5\over 2}, \gamma^\mu, \gamma^\mu\gamma_5, \sigma^{\mu\nu}\}$ or equals $(\bar q\gamma^\mu q)$ for the non-local dim-5 case.

The neutrino current $L_{\rm inv}$ can be easily identified for a given specific effective dim-5/dim-6 interaction. For the calculation of ${\cal H}^{\Gamma}$, we start from the matrix element of the quark-level process $\bar q(p_{\bar q})q(p_q)\to\gamma(k){\rm inv}_1(k_1){\rm inv}_2(k_2)$. The quark-level transition $h^\Gamma=\langle\gamma(k)|(\bar q\Gamma q)|\bar q(p_{\bar q})q(p_q)\rangle$, based on the diagrams in Fig.~\ref{Fig1}, is given by
\begin{eqnarray}\nonumber
h^\Gamma&=&\bar v_{\bar q} \left[\left(-ieQ_q\slashed{\epsilon}_\gamma^*\right){i\over \slashed{k}-\slashed{p}_{\bar q}-m_q}\Gamma+\Gamma{i\over \slashed{p}_{q}-\slashed{k}-m_q}\left(-ieQ_q\slashed{\epsilon}_\gamma^*\right)\right]u_q
\\\nonumber
&=&eQ_q \bar v_{\bar q} \left[\slashed{\epsilon}_\gamma^*{\slashed{k}-\slashed{p}_{\bar q}+m_q\over (k-p_{\bar q})^2-m_q^2}\Gamma+\Gamma{\slashed{p}_{q}-\slashed{k}+m_q\over (p_{q}-k)^2-m_q^2}\slashed{\epsilon}_\gamma^*\right]u_q
\\
&\overset{\rm on-shell}{=}&eQ_q \bar v_{\bar q} \left[{2p_{\bar q}\cdot\epsilon_\gamma^*-\slashed{\epsilon}_\gamma^*\slashed{k}\over 2p_{\bar q}\cdot k}\Gamma-\Gamma{2p_{q}\cdot\epsilon_\gamma^* -\slashed{k}\slashed{\epsilon}_\gamma^*\over 2p_{q}\cdot k}\right]u_q\;.
\label{amp1}
\end{eqnarray}
For the further reduction of the above amplitude, we work in the rest frame of $J/\psi$ state and use the non-relativistic color singlet model (NRCSM) to calculate the hadronic matrix element~\cite{Appelquist:1974zd,DeRujula:1974rkb,Kuhn:1979bb,Keung:1980ev,Berger:1980ni,Clavelli:1982hp,Clavelli:2001zi,Clavelli:2001gb}. In the NRCSM, the charm quark pair $c$ and $\bar c$ within $J/\psi$ is treated as static constituent quarks and the momentum (mass) of $c/\bar c$ is taken to be one-half of the momentum (mass) of the state $J/\psi$, i.e. $q_c=q_{\bar c}=P/2~(m_c=m_{\bar c}=m_{J}/2)$~\footnote{For brevity, throughout the whole context, we abbreviate $X_{J/\psi}$ to $X_J$ to specify the $X$ property of $J/\psi$ state.}~\footnote{$m_c^{\rm PDG}=1.27\pm0.02{\rm~GeV}, m_J^{\rm PDG}=3096.9\pm0.006\rm~MeV$.}. Thus, the difference between $m_J/2$ and $m_c$ scales as ${\cal O}(\Lambda_{\rm QCD})$ and can be neglected together with the quark-antiquark Fermi motion effects. The $J/\psi$ state is formed from the quark-antiquark pair $c\bar c$ through the projection operation~\cite{Hao:2006nf}
\begin{align}
\bar v_{\bar c}\Gamma u_{c}\to {\Psi(0)N_c\over \sqrt{12m_J}}\Tr\left[(\slashed{P}+m_J)\slashed{\epsilon}_J\Gamma \right]\;,
\end{align}
where $\epsilon_J^\mu$ is the polarization vector of $J/\psi$, $N_c=3$ is the color factor in the fundamental representation of SU$(3)_c$ and the wave-function of $J/\psi$ at origin is denoted by $\Psi(0)$. Given the above assumptions, Eq.~\eqref{amp1} is translated to the hadronic matrix element ${\cal H}^\Gamma$
\begin{eqnarray}\nonumber
h^\Gamma&=&{eQ_c\over  P\cdot k} \bar v_{\bar c} \left[\left(P\cdot\epsilon_\gamma^*-\slashed{\epsilon}_\gamma^*\slashed{k}\right)\Gamma-\Gamma\left(P\cdot\epsilon_\gamma^* -\slashed{k}\slashed{\epsilon}_\gamma^*\right)\right]u_c
\\
&\to&{\cal H}^\Gamma={eQ_c\over  P\cdot k} {\Psi(0)N_c\over \sqrt{12m_J}} \Tr\left[(\slashed{P}+m_J)\slashed{\epsilon}_J\{\slashed{k}\slashed{\epsilon}_\gamma^*,\Gamma\}\right]\nonumber
\\
&&~~~~~={\cal N}(q^2)\times
 \begin{cases}
\left[ P^\sigma k^\rho- g^{\rho\sigma}(P\cdot k)\pm i\epsilon^{\mu\nu\rho\sigma}P_\mu k_\nu \right]\epsilon_{J,\rho}\epsilon_{\gamma,\sigma}^*\;, &\Gamma=P_\pm
\\
-2 im_J\epsilon^{\mu\nu\rho\sigma}k_\nu\epsilon_{J,\rho}\epsilon_{\gamma,\sigma}^*\;, &
\Gamma=\gamma^\mu \gamma_{5}
\\
0\;, &\Gamma=\gamma^\mu\;, \sigma^{\mu\nu}
\end{cases}
\end{eqnarray}
where the momentum $q=P-k$ and
\begin{align}
{\cal N}(q^2)={4eQ_c\over  P\cdot k} {\Psi(0)N_c\over \sqrt{12m_J}}
= {2\sqrt{3}eQ_c\Psi(0)\over \sqrt{m_J}}{1\over  P\cdot k}\;.
\end{align}
On the other hand, the non-vanishing hadronic matrix elements for the charmonium meson $J/\psi$ with momentum $P$ and polarization vector $\epsilon^\mu_J$ can also be parameterized as~\cite{Ball:2006eu,Cheng:2013fba}
\begin{align}
\langle 0|\bar{c}\gamma^\mu c|J/\psi(P)\rangle =& f_J m_J \epsilon^\mu_J \; , &
\langle 0|\bar{c}\sigma^{\mu\nu} c|J/\psi(P)\rangle =& if_J^{T} \left(\epsilon^\mu_J P^\nu - \epsilon^\nu_JP^\mu \right) \; ,
\label{formfactor}
\end{align}
where $f_J$ and $f_J^T$ are the vector and tensor decay constants for the vector meson $J/\psi$, and $m_J$ is the mass of $J/\psi$.
Converting the general parameterization in terms of decay constants into the NRCSM formalism, we have the following relationship
\begin{align}
&f_J={4\Psi(0)N_c\over \sqrt{12 m_J}}= {2\sqrt{3}\Psi(0)\over \sqrt{m_J}}\;,
&&f_J^T=f_J\;.
\end{align}
In our numerical calculation below, we adopt the NRCSM formalism and determine the wave-function in terms of the branching
fraction to $e^+e^-$
\begin{align}
{\cal B}(J/\psi\to e^+e^-)={16\pi Q_c^2\alpha_{\rm em}^2 |\Psi_J(0)|^2 \over m_J^2\Gamma_J}
={4\pi Q_c^2 \alpha_{\rm em}^2f_J^2 \over 3 m_J\Gamma_J}\;.
\end{align}

From the above results, we conclude that the contribution to the radiative transition $J/\psi\to \gamma+\rm inv_1 + inv_2$ from the operators with a pure vector current (the case of dim-5 dipole operators) or a tensor current (the case of dim-6 tensor quark current) vanishes. This is understandable due to the charge conjugation symmetry of QCD and QED since $J/\psi$, the photon and the pure vector/tensor current all have negative charge parity. Thus, only the Lorentz structures of $P_\pm$ and $\gamma^\mu P_\pm$ are to be considered below. For a three-body decay $V\to a+b+c$, the decay width becomes
\begin{eqnarray}
{d\Gamma\over dx_a dx_b}={m_V\over 256\pi^3} \overline{|\mathcal{M}|^2}\;,
\end{eqnarray}
where $x_i=2E_i/m_V~(i=a,b,c)$ and $x_a+x_b+x_c=2$.
The kinematics constrains the domain of $x_a$ and $x_b$ to be~\cite{CP}
\begin{eqnarray}
2\mu_a^{1/2}&\leq& x_a \leq 1+\mu_a - \mu_b - \mu_c - 2(\mu_b \mu_c)^{1/2}\;,\nonumber\\
x_{b, {\rm max(min)}}&=&{1\over 2}(1-x_a+\mu_a)^{-1} \Big[(2-x_a)(1+\mu_a + \mu_b - \mu_c - x_a)\nonumber \\
&&\pm (x_a^2-4\mu_a)^{1/2}\lambda^{1/2}(1+\mu_a-x_a,\mu_b,\mu_c)\Big] \;,
\label{kin}
\end{eqnarray}
where
\begin{eqnarray}
&&\mu_i=m_i^2/m_V^2\;, \ i=a,b,c\;,\nonumber\\
&&\lambda(x,y,z)=x^2+y^2+z^2-2xy-2yz-2xz\;.
\end{eqnarray}
In our case we take $a=\gamma$ and $b,c=\nu, N$. The differential decay width against the photon energy $E_\gamma$ is
\begin{align}
{d\Gamma\over dE_\gamma}={2\over m_J}{d\Gamma\over dx_\gamma}
={1\over 128\pi^3}\int d x_b\overline{|{\cal M}|^2}\;.
\end{align}
We show the matrix elements of $J/\psi$ decay in LNEFT and the kinematic functions in Appendix~\ref{sec:ME-LNEFT}.
The non-vanishing partial widths are governed by $\mathcal{O}_{c\nu 1,2}^V$, $\mathcal{O}_{cN 1,2}^V$, $\mathcal{O}_{c\nu N 1,2}^S$ in LNC case or $\mathcal{O}_{c\nu 1,2}^S$, $\mathcal{O}_{cN 1,2}^S$, $\mathcal{O}_{c\nu N 1,2}^V$ in LNV case.
In particular, in the SM the vector current operator are generated with the Wilson coefficients $C_{q \nu 1(2), \mathrm{SM}}^{V, p r \alpha \beta}=-2\sqrt{2}G_F\left(T_{3}-Q_{q} s_{W}^{2}\right) \delta_{p r} \delta_{\alpha \beta}$. The SM amplitude becomes
\begin{align}
{\cal M}_{\rm SM}=-i4\sqrt{6}eQ_cG_F{\sqrt{m_J}\Psi(0)\over  m_J^2-q^2}\epsilon^{\mu\nu\rho\sigma}k_\nu\epsilon_{J,\rho}\epsilon_{\gamma,\sigma}^* (\bar u_{\nu}\gamma^\mu P_Lv_{\bar \nu})\;.
\end{align}

The above amplitude leads to the differential decay width being
\begin{align}
{d\Gamma\over dE_\gamma}\Big|_{\rm SM}={4Q_c^2\alpha_{\rm em} |\Psi(0)|^2G_F^2 \over 3\pi^2m_J}E_\gamma(m_J-E_\gamma)\;,
\Gamma_{\rm SM}&=N_\nu \int dE_\gamma{d\Gamma\over dE_\gamma}\Big|_{\rm SM}
={Q_c^2 \alpha_{\rm em} |\Psi(0)|^2G_F^2m_J^2 \over 3\pi^2}\;,
\end{align}
where $N_\nu=3$ is the number of active neutrinos.
Our SM result agrees with that in Ref.~\cite{Gao:2014yga}.
In the numerical analysis below, we work on the case of one flavor sterile neutrino and denote its mass as $m_N$.
For the operators involving active neutrinos, we assume single-flavor dominance for the flavor dependent couplings.

\subsection{The radiative decay of heavy quarkonium in DMEFT}
\label{sec:decayinDMEFT}

For the radiative decay of $J/\psi$ into DM pairs, analogous to the neutrino case, the dim-5 dipole operators ${\cal O}_{\chi F1,2}$, the dim-6/7 vector quark current operators ${\cal O}_{\chi q1,2}, {\cal O}_{\chi q11,12}$ and the dim-7 tensor quark operators ${\cal O}_{\chi q9,10}, \hat{\cal O}_{\chi q15,16,17,18}$ have vanishing contributions. Besides the diagrams in Fig.~\ref{Fig1}, the photon may also be emitted from the interacting vertex induced by the Rayleigh operators in DMEFT. However, the Rayleigh operators do not contribute to the $J/\psi$ radiative decay either due to the nature of vector quark current from the intermediate photon.

The non-vanishing matrix elements for the process $J/\psi(P)\to\gamma(k)\chi(k_1)\bar\chi(k_2)$ from the remaining operators ${\cal O}_{\chi q3,4,5,6,7,8,13,14}$ are as follows
\begin{align}\nonumber
{{\cal M}_{\chi\bar\chi}\over m_J{\cal N}(q^2)}&=-2i\epsilon^{\mu\nu\rho\sigma}k_\nu\epsilon_{J,\rho}\epsilon^*_{\gamma,\sigma}\overline{u_\chi}\left[\gamma_\mu(C_{\chi c3}+  C_{\chi c4}\gamma_5) +(k_1-k_2)_\mu(C_{\chi c13}+C_{\chi c14}i\gamma_5  )\right]v_{\bar\chi}
\\\nonumber
&+\left(P^\sigma k^\rho-g^{\rho\sigma}(P\cdot k)\right)\epsilon_{J,\rho}\epsilon^*_{\gamma,\sigma} \overline{u_\chi}\left[C_{\chi c5}+  C_{\chi c6}i\gamma_5\right]v_{\bar\chi}
\\
&-\epsilon^{\mu\nu\rho\sigma}P_\mu k_\nu\epsilon_{J,\rho}\epsilon^*_{\gamma,\sigma}  \overline{u_\chi}\left[C_{\chi c7}+  C_{\chi c8}i\gamma_5\right]v_{\bar\chi}\;,
\end{align}
where we have rewritten
$2m_c$ as $m_J$ in the second and third lines. The squared and spin-summed matrix element is
\begin{align}
\overline{|{{\cal M}_{\chi\bar\chi}}|}^2=16e^2Q_c^2|\Psi_J(0)|^2m_J{\cal R}\;,
\end{align}
where ${\cal R}$ is defined as
\begin{align}\nonumber
{\cal R}&=2\left[8g(0)-f(8\mu_\chi)\right]C_{\chi c3}^2+2\left[8g(0)-f(-4\mu_\chi)\right]C_{\chi c4}^2
\\\nonumber
&+m_J^2\left[f(4\mu_\chi)\left(C_{\chi c5}^2+2C_{\chi c7}^2+4h(\mu_\chi)C_{\chi c13}^2\right)+  f(0)\left(C_{\chi c6}^2+2C_{\chi c8}^2+4h(\mu_\chi)C_{\chi c14}^2\right)\right]
\\
&-16m_J\left[\sqrt{\mu_\chi} h(\mu_\chi)C_{\chi c3}C_{\chi c13}-k(\mu_\chi)C_{\chi c4}C_{\chi c8}\right]\;,
\end{align}
with $\mu_\chi=m^2_\chi/m_J^2$ and
\begin{align}
h(\mu_\chi)={1\over x_\gamma^2}\left[(x_b-x_c)^2(1+x_\gamma)+(1-x_\gamma-4\mu_\chi)x_\gamma^2\right]\;,\;
k(\mu_\chi)={\sqrt{\mu_\chi}\over x_\gamma}(2-x_b-x_c)\;.
\end{align}
The $f$ and $g$ functions are defined in Eq.~(\ref{fgfunction}) in Appendix~\ref{sec:ME-LNEFT}.
We then arrive at the final expression of
the differential partial width
\begin{align}\nonumber
{d\Gamma\over dE_\gamma}&={16e^2Q_c^2|\Psi_J(0)|^2m_J\over 128\pi^3}\times {x_\gamma\sqrt{(1-x_\gamma)(1-x_\gamma-4\mu_\chi)}\over 3(1-x_\gamma^2)}
\\\nonumber
&
\times\left[ 2\left(13-18x_\gamma+5x_\gamma^2+16\mu_\chi(2-x_\gamma)\right)C_{\chi c3}^2
+2\left(13-18x_\gamma+5x_\gamma^2-4\mu_\chi(1-5x_\gamma)\right)C_{\chi c4}^2
\right.
\\\nonumber
&\left.
+3m_J^2(1-x_\gamma)(1-x_\gamma-4\mu_\chi)\left(C_{\chi c5}^2+2C_{\chi c7}^2 \right)
+3m_J^2(1-x_\gamma)^2\left(C_{\chi c6}^2+2C_{\chi c8}^2 \right)
\right.
\\\nonumber
&\left.
+8m_J^2(2-x_\gamma)(1-x_\gamma-4\mu_\chi)^2C_{\chi c13}^2
+8m_J^2(2-x_\gamma)(1-x_\gamma)(1-x_\gamma-4\mu_\chi)C_{\chi c14}^2
\right.
\\
&\left.
-32m_J\sqrt{\mu_\chi}(2-x_\gamma)(1-x_\gamma-4\mu_\chi)C_{\chi c3}C_{\chi c13}
+48m_J\sqrt{\mu_\chi}(1-x_\gamma)C_{\chi c4}C_{\chi c8}
\right]\;.
\end{align}
In the following numerical analysis, we
assume the dominance of one Wilson coefficient at a time.

\section{The heavy quarkonium decay into invisible particles}
\label{sec:invdecay}
From the decay constants in Eq.~(\ref{formfactor}) and the fact that $J/\psi$ is a vector meson with $J^{PC}=1^{--}$, only the dipole operators and the dim-6/7 operators with a vector or tensor quark current will non-trivially contribute to the invisible decay $J/\psi\to \rm inv_{1}+inv_{2}$.
In particular, the dipole operators contribute to the invisible decay through a photon propagator and a QED vertex.

\subsection{The invisible decay of heavy quarkonium in LNEFT}
\label{sec:invdecayinLNEFT}

For $J/\psi(P)\to {\rm inv}_{1/\alpha}(k_{1})+{\rm inv}_{2/\beta}(k_{2})$, after squaring the amplitudes and including the phase space factor, the final branching ratio is given by~\cite{Li:2020lba}
\begin{align}\nonumber
{\cal B}(J/\psi\to {\rm inv.})
&=\frac{m_J^3}{48\pi \Gamma_{J}}\sum_{\alpha,\beta}\Bigg\{
2\left|\frac{f_J}{2}\left( C_{c\nu 1, {\rm NP}}^{V,\alpha\beta}+C_{c\nu 2, {\rm NP}}^{V,\alpha\beta}+2 C^V_{\rm SM}\delta_{\alpha\beta}\right)\right|^2
\\\nonumber
&+2\left|\frac{f_J}{2}\left( C_{cN1}^{V,\alpha\beta}+C_{cN 2}^{V,\alpha\beta}\right)\right|^2
\left(1-\frac{m_N^2}{m_J^2}\right)\left(1-4\frac{m_N^2}{m_J^2}\right)^{1\over 2}
\\\nonumber
&+8\left|f_J^{T}C_{c\nu N}^{T,\alpha\beta} -eQ_c{f_J\over m_J}C_{\nu NF}^{\alpha\beta}\right|^2
\left(1+{m_N^2 \over m_J^2}-2{m_N^4 \over m_J^4}\right)\left(1-\frac{m_N^2}{m_J^2}\right)
\\\nonumber
&+16\left|f_J^{T}C_{c\nu}^{T,\alpha\beta}-eQ_c{f_J\over m_J}C_{\nu\nu F}^{\alpha\beta}\right|^2
\\\nonumber
&+16\left|f_J^{T}C_{cN}^{T,\alpha\beta}-eQ_c{f_J\over m_J}C_{NN F}^{\alpha\beta}\right|^2
\left(1+2{m_N^2\over m_J^2}\right)\left(1-4{m_N^2\over m_J^2}\right)^{1\over2}
\\
&+4\left|{f_J\over 2}\left(C_{c\nu N1}^{V,\alpha\beta}+ C_{c\nu N2}^{V,\alpha\beta}\right)\right|^2
\left(1-{m_N^2 \over 2m_J^2}-{m_N^4 \over 2m_J^4}\right)\left(1-{m_N^2\over m_J^2}\right)
\Bigg\}\;.
\label{eq:BrVinv}
\end{align}
The non-vanishing partial widths are determined by $\mathcal{O}_{\nu NF}$, $\mathcal{O}_{c\nu 1,2}^V$, $\mathcal{O}_{c N 1,2}^V$, $\mathcal{O}_{c\nu N}^T$ in LNC case or $\mathcal{O}_{\nu\nu F}$, $\mathcal{O}_{NNF}$, $\mathcal{O}_{c\nu N 1,2}^V$, $\mathcal{O}_{c\nu}^T$, $\mathcal{O}_{cN}^T$ in LNV case.
Here we have split the contribution from the vector operators $C_{c\nu 1(2)}^V$ into the NP contribution and the SM part with
$C^V_{\rm SM}=-{g_Z^2\over4m_Z^2}\Big({1\over2}-{4\over3}s_W^2\Big)$.
Taking the SM Wilson coefficient $C_{\rm SM}^{V}$, we obtain the same analytic expression of the ratio $\tfrac{\Gamma(J/\psi\to \nu\bar{\nu})}{\Gamma(J/\psi\to e^{+}e^{-})} $ as that in Ref.~\cite{Chang:1997tq}. Our numerical SM prediction gives $3.23\times 10^{-7}$ for the above ratio and ${\cal B}(J/\psi\to \nu\bar{\nu})=1.93\times 10^{-8}$ given ${\cal B}(J/\psi\to e^{+}e^{-})=(5.971\pm0.031)\%$~\cite{Zyla:2020zbs}. This is compatible with the estimate in Ref.~\cite{Fayet:1979qi}, but smaller than the value in Ref.~\cite{Chang:1997tq} by 29\%.

\subsection{The invisible decay of heavy quarkonium in DMEFT}
\label{sec:invdecayinDMEFT}

Analogously, the invisible decay of $J/\psi$ in the DMEFT also receives non-vanishing contributions from the operators with a quark vector or tensor current and the dipole operators.
The amplitude for the process $J/\psi\to \chi(k_{1})\bar\chi(k_{2})$ from the interactions in Eqs.~(\ref{eq: dmF}-\ref{eq: dmq}) can be written as ${\cal M}(J/\psi\to \chi\bar\chi)={\cal M}_{\mu}\epsilon_{J}^{\mu}$ with
\begin{align}\nonumber
{\cal M}_{\mu}&=m_{J}f_{J}\overline{u_{\chi}}\Big[
-2ieQ_{c}{P^{\nu}\over m_{J}^{2}}\sigma_{\mu\nu}\left(C_{\chi F1}+C_{\chi F2}i\gamma_{5}\right)
+\gamma_{\mu}(C_{\chi c1}+C_{\chi c2}\gamma_{5})
\\\nonumber
&
\quad\quad\quad\quad\quad\; +(k_{1}-k_{2})_{\mu}\left(C_{\chi c11}+C_{\chi c12}i\gamma_{5}\right)\Big]v_{\bar\chi}
\\\nonumber
&+2if_{J}^{T}P^{\nu}\overline{u_{\chi}}\Big[
m_{c}\sigma_{\mu\nu}(C_{\chi c9}+C_{\chi c10}i\gamma_{5})
+\gamma_{[\mu}(k_{1}-k_{2})_{\nu]}(C_{\chi c15}+C_{\chi c16}\gamma_{5}) \Big]v_{\bar\chi}
\\
&-i\epsilon_{\mu\nu\rho\sigma}f_{J}^{T}P^{\nu}\overline{u_{\chi}}\Big[
\gamma^{[\rho}(k_{1}-k_{2})^{\sigma]}(C_{\chi c17}+C_{\chi c18}\gamma_{5})\Big]v_{\bar\chi}\;,
\end{align}
where one can see that the term proportional to $C_{\chi c15}$ vanishes once the on-shell condition applies.
We then obtain the branching ratio as
\begin{align}
{\cal B}(J/\psi\to \chi\bar\chi)&={1\over \Gamma_{J}}{1\over 16\pi m_{J}}(1-4\mu_{\chi})^{1\over 2}\overline{\left| {\cal M}(J/\psi\to \chi\bar\chi) \right|}^{2}
\\\nonumber
&=\frac{m_J^3}{48\pi \Gamma_{J}}(1-4\mu_{\chi})^{1\over 2}\Big[
8e^{2}Q_{c}^{2}{f_{J}^{2}\over m_{J}^{2}} \left[ (1+8\mu_{\chi})C_{\chi F1}^{2}+ (1-4\mu_{\chi})C_{\chi F2}^{2} \right]
\\\nonumber
&+4f_{J}^{2} \left[(1+2\mu_{\chi})C_{\chi c1}^{2}+ (1-4\mu_{\chi})C_{\chi c2}^{2} \right]
\\\nonumber
&+2\left(f_{J}^{T}\right)^{2}\left [(1+8\mu_{\chi}) C_{\chi c9}^{2}+ (1-4\mu_{\chi})C_{\chi c10}^{2} \right]
\\\nonumber
&+2m_{J}^{2} f_{J}^{2}(1-4\mu_{\chi})\left[(1-4\mu_{\chi})C_{\chi c11}^{2}+ C_{\chi c12}^{2}\right]
\\\nonumber
&+16m_{J}^{2}\left(f_{J}^{T}\right)^{2}(1-4\mu_{\chi})\left[2\mu_{\chi}C_{\chi c16}^{2}+ C_{\chi c17}^{2}+ (1-4\mu_{\chi})C_{\chi c18}^{2}\right]
\Big]+\cdots\;,
\end{align}
where $\mu_{\chi}=m_{\chi}^{2}/m_{J}^{2}$ as defined before and $\cdots$ stands for the interference terms which are omitted for simplicity.
The above terms are from the operators which have vanishing contributions to the $J/\psi\to \gamma+{\rm invisible}$ process.

\section{Numerical results}
\label{sec:Num}

\begin{figure}
\includegraphics[width=0.48\linewidth]{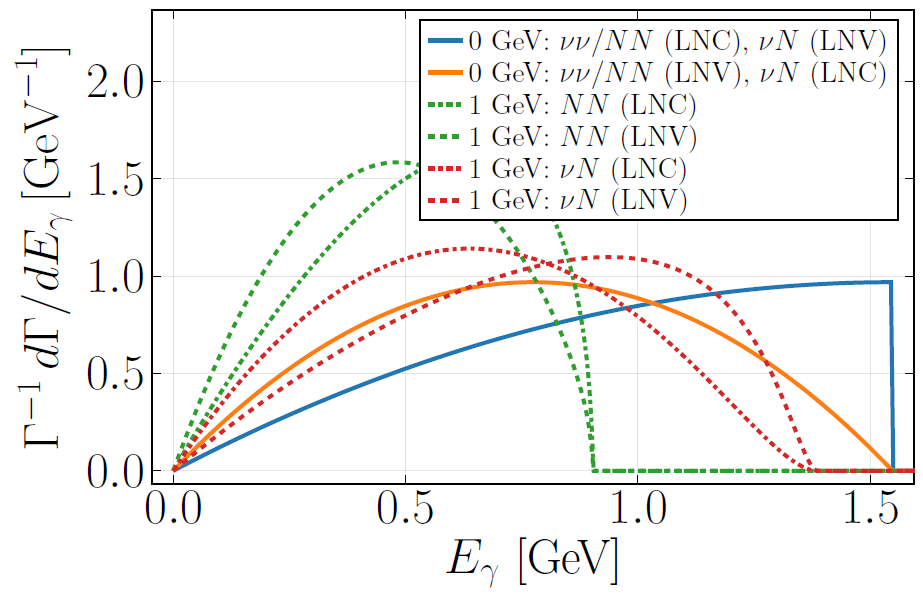}
\includegraphics[width=0.48\linewidth]{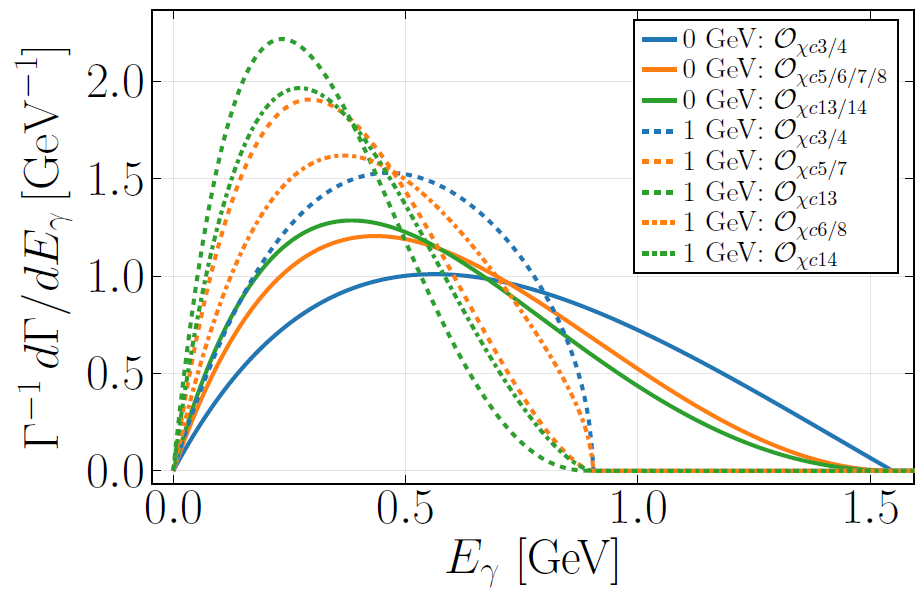}
\caption{Normalized differential width distributions as a function of photon energy $E_\gamma$ for $J/\psi\to \gamma+{\rm invisible}$ in LNEFT (left) and DMEFT (right). In the left panel, massless sterile neutrinos are shown with solid lines and sterile neutrinos with masses $m_N=1$ GeV with dashed and dot-dashed lines. Blue corresponds to the channels $\nu\nu$ and $NN$ ($\nu N$) for LNC (LNV) interactions. Orange corresponds to the channels $\nu\nu$ and $NN$ ($\nu N$) for LNV (LNC) interactions. Green (red) stands for the $NN$ ($\nu N$) channel, while dashed (dot-dashed) lines refer to LNV (LNC) interactions.
In the right panel, solid lines refer to massless dark matter, and dashed and dot-dashed lines show the case of $m_\chi=1$ GeV. Blue (orange) [green] lines refer to the operators $\mathcal{O}_{\chi c3/4}$ ($\mathcal{O}_{\chi c 5/6/7/8}$) [$\mathcal{O}_{\chi c 13/14}$].}
\label{dis}
\end{figure}

In this section we show the constraints from both $J/\psi\to \gamma+{\rm invisible}$ and $J/\psi\to {\rm invisible}$.
We first show the normalized differential width distributions for $J/\psi\to \gamma+{\rm invisible}$ in LNEFT (left) and DMEFT (right) in Fig.~\ref{dis} as a function of the photon energy $E_\gamma$.
One can see that the distributions are determined by both the individual Lorentz structure and the mass of sterile neutrino $N$ or DM particle $\chi$.
When the mass of sterile neutrino is negligible, the LNC $\nu\bar{\nu}, N\bar{N}$ ($\nu\bar{N}/\bar{\nu}N$) cases and the LNV $\nu N/\bar{\nu}\bar{N}$ ($\nu \nu/\bar{\nu}\bar{\nu}, NN/\bar{N}\bar{N}$) cases share the same $E_\gamma$ distribution. The maximal value of $E_\gamma$ depends on $m_N$ and how many massive sterile neutrinos are produced in $J/\psi$ decay. For negligible DM mass in $\mu_\chi=m^2_\chi/m_J^2$, the distributions of $\mathcal{O}_{c\chi 3}$ ($\mathcal{O}_{c\chi 5,7}$) [$\mathcal{O}_{c\chi 13}$] and $\mathcal{O}_{c\chi 4}$ ($\mathcal{O}_{c\chi 6,8}$) [$\mathcal{O}_{c\chi 14}$] are the same. Based on Eq.~(\ref{kin}) and the minimal photon energy as 1.25 GeV, the BESIII data can constrain $m_N$ up to 1.36 (0.68) GeV for one (two) sterile neutrino in the final states of $J/\psi$ decay. As the DM particle is always produced in pairs, the BESIII data constrain the DM mass to 0.68 GeV at most.

\subsection{Fit to the experimental data of $J/\psi\to \gamma+{\rm invisible}$}

In our fit to the experimental data we follow as closely as possible the experimental analysis~\cite{Ablikim:2020qtn}. We assume the events in each bin to be distributed following a Poisson distribution and thus the likelihood is given by
\begin{equation}
  \mathcal{L} = \prod_{i=1}^{N_{\rm bins}} P\left(N_i\Bigg|\mathcal{B}\,\epsilon_{\rm sig}\, s_i\, N_{J/\psi}/\epsilon_{J/\psi} + \sum_{j}^{N_{\rm bkg}} b_{ij}^{\rm exp}\right)\;,
\end{equation}
where $N_{\rm bins}$ denotes the number of bins, $N_i$ the number of events in the $i$-th bin, $\mathcal{B}$ the branching ratio of $J/\psi\to \gamma + \rm
inv$, $\epsilon_{\rm sig}=0.93$ is the signal efficiency which is
characterized by the acceptance of photons in the detector,
$N_{J/\psi}=(8848\pm1)\times 10^4$ is the number of tagged $J/\psi$ events in
the signal region, $\epsilon_{J/\psi}=0.5680\pm 0.0001$ denotes the
efficiency for tagging a $J/\psi$, $s_i$ represents the signal probability in
the $i$-th bin and $b_{ij}^{\rm exp}$ denotes the number of background events
of type $j$ in the $i$-th bin. The number of background events in each bin
have been extracted from Fig.~\ref{dis} in Ref.~\cite{Ablikim:2020qtn}. The signal
probability is given by integrating the partial width for the process of
interest for each bin and normalizing it to the total partial width, i.e.
$s_i=\Gamma^{-1}\int_{\rm bin\, i} dE_\gamma d\Gamma/dE_\gamma$. Systematic
uncertainties are taking into account using Gaussian distributions for the
nuisance parameters, $N_{J/\psi}$, $\epsilon_{J/\psi}$, $\epsilon_{\rm sig}$
and the normalizations $N_{J/\psi\to\pi^0\gamma}$, $N_{J/\psi\to \eta \gamma}$, $N_{\rm cont}$ of the different background distributions from $J/\psi\to \pi^0\gamma$, $J/\psi\to \eta\gamma$ and the continuum background, respectively. The central values for the normalizations $N_{J/\psi\to\pi^0\gamma}$, $N_{J/\psi\to \eta\gamma}$ and $N_{\rm cont}$ are taken to be unity.
For the standard deviations of the Gaussian distributions, we use
$\sigma_{N_{J/\psi}}=0.007 N_{J/\psi}$, $\sigma_{\epsilon_{J/\psi}}=0.0001$,
$\sigma_{\epsilon_{\rm sig}}=0.015 \epsilon_{\rm sig}$, $\sigma_{J/\psi\to
\pi^0\gamma}=0.17 N_{J/\psi\to \pi^0\gamma}$, $\sigma_{\rm J/\psi\to
\eta\gamma}=0.17 N_{J/\psi\to \eta\gamma}$, and $\sigma_{\rm cont}=0.044 N_{\rm cont}$.

Following the experimental analysis~\cite{Ablikim:2020qtn}, we set limits using the CL$_s$ method~\cite{Read:2000ru,Read:2002hq} with the profile likelihood ratio as test statistic. For the calculation we use the approximation based on the Asimov dataset detailed in Ref.~\cite{Cowan:2010js}. We validated our implementation of the fit calculation by reproducing the experimental limits for the 2-body decays studied in Ref.~\cite{Ablikim:2020qtn}. The constraints for the 3-body final states are generally weaker due to the broader differential width distribution and the lower photon energy cutoff in the experimental analysis.

\subsection{Constraints on LNEFT and DMEFT }
\label{sec:limit}

\begin{figure}
\includegraphics[width=0.48\linewidth]{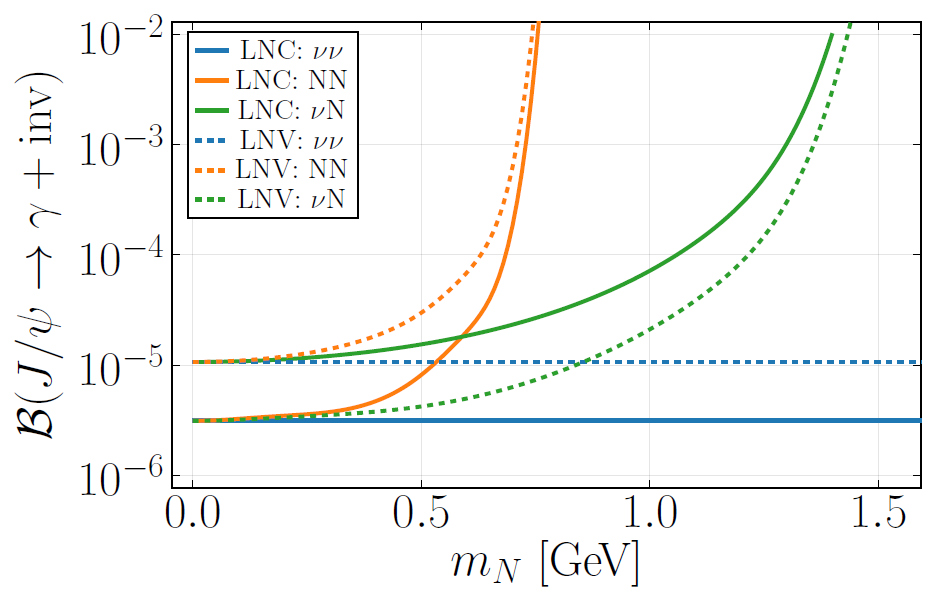}
\includegraphics[width=0.48\linewidth]{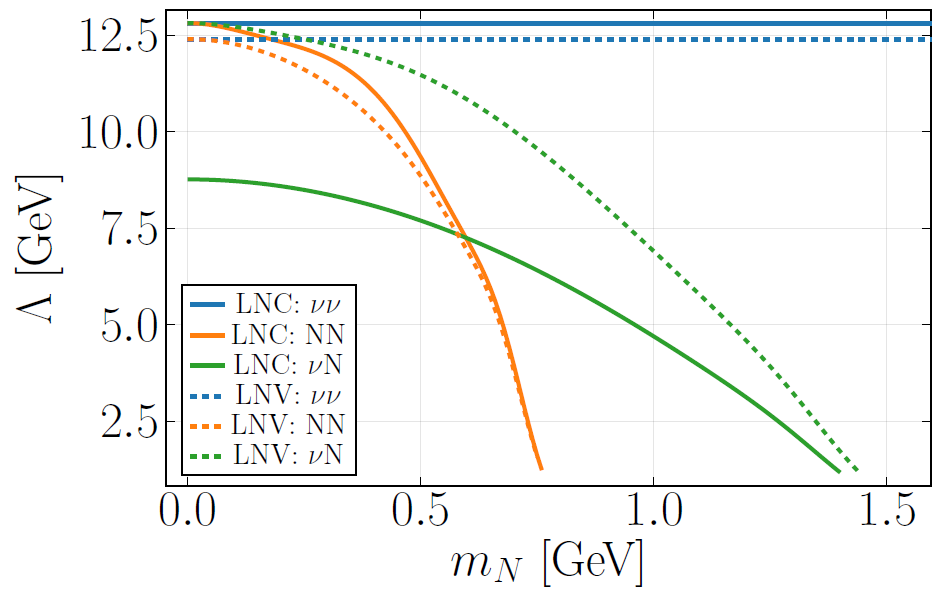}\\
\includegraphics[width=0.49\linewidth]{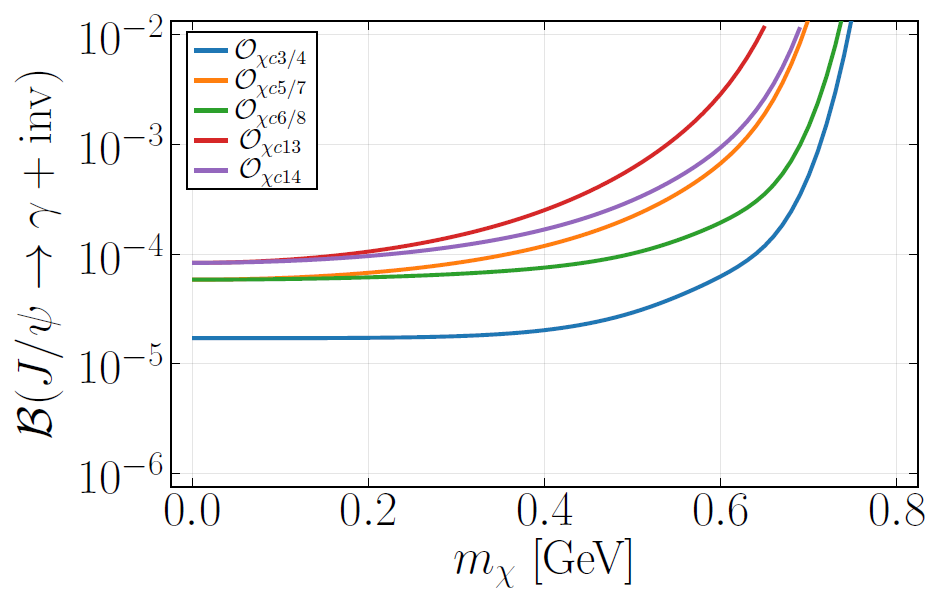}
\includegraphics[width=0.49\linewidth]{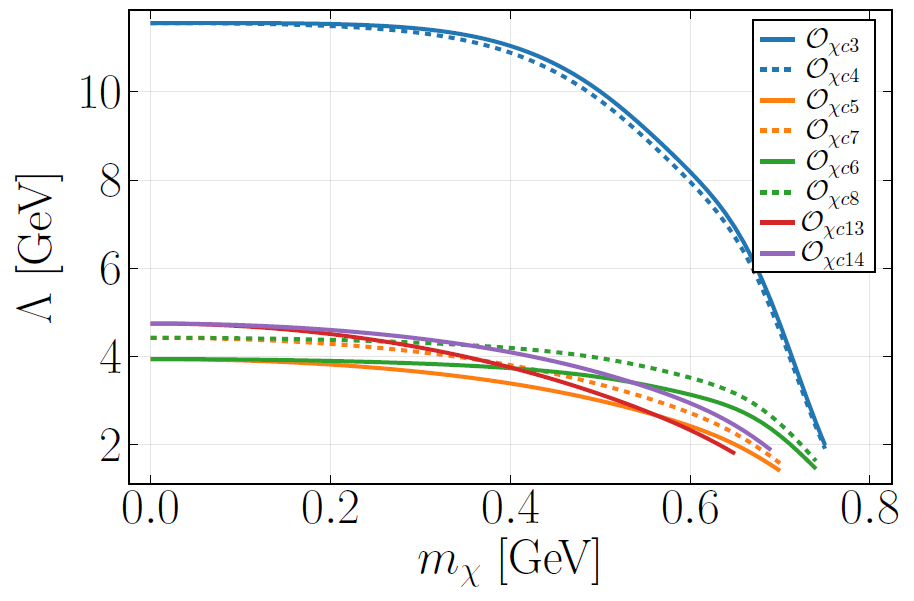}
\caption{Constraints on different neutrino scenarios in LNEFT (top) and DMEFT (bottom) from $J/\psi\to \gamma+{\rm invisible}$. Left: upper limits on the decay branching fractions. Right: lower limits on the energy scales. For the LNEFT figures on the top, solid (dashed) lines represent LNC (LNV) interactions. Blue (orange) [green] lines correspond to the scenarios $\nu\nu$ ($NN$) [$\nu N$]. For the DMEFT figures on the bottom, blue, orange, green, red, purple corresponds to operators $\mathcal{O}_{\chi c 3/4}$, $\mathcal{O}_{\chi c 5/7}$, $\mathcal{O}_{\chi c 6/8}$, $\mathcal{O}_{\chi c 13}$, $\mathcal{O}_{\chi c 14}$. In the bottom right panel, solid lines correspond to operators $\mathcal{O}_{\chi c 3/5/6/13/14}$ while dashed lines correspond to operators $\mathcal{O}_{\chi c 4/7/8}$.
}
\label{fig:radlimit}
\end{figure}

The upper panels of Fig.~\ref{fig:radlimit} show the upper limits on the decay branching fractions of different neutrino scenarios as a function of sterile neutrino mass from $J/\psi\to \gamma+{\rm invisible}$. For massless sterile neutrino, the limits on the branching fractions of LNC $\nu\nu, NN$, LNV $\nu N$ and LNC $\nu N$, LNV $\nu\nu, NN$ cases are $3\times 10^{-6}$ and $10^{-5}$, respectively.
One can then convert the decay branching fraction bounds into the lower limits on the energy scale associated with the corresponding Wilson coefficients. The most stringent bound on the energy scale is 12.8 GeV for the LNC operators $\mathcal{O}_{c\nu 1,2}^V$.

For different dark matter scenarios in DMEFT, the lower panels of Fig.~\ref{fig:radlimit} show the upper limits on the decay branching fractions as a function of $m_\chi$ from $J/\psi\to \gamma+{\rm invisible}$. Due to the suppression of $\mu_\chi$ for very small $m_\chi$, the BR limits for $\mathcal{O}_{c\chi 3}$ ($\mathcal{O}_{c\chi 5,7}$) [$\mathcal{O}_{c\chi 13}$] and $\mathcal{O}_{c\chi 4}$ ($\mathcal{O}_{c\chi 6,8}$) [$\mathcal{O}_{c\chi 14}$] are equal and become $2\times 10^{-5}$ ($6\times 10^{-5}$) [$8\times 10^{-5}$]. As a result, as shown in Fig.~\ref{fig:radlimit} (bottom right), the most stringent bound on the energy scale is 11.6 GeV for the operators $\mathcal{O}_{c\chi 3,4}$.

\begin{figure}
\centering
\includegraphics[width=7.5cm]{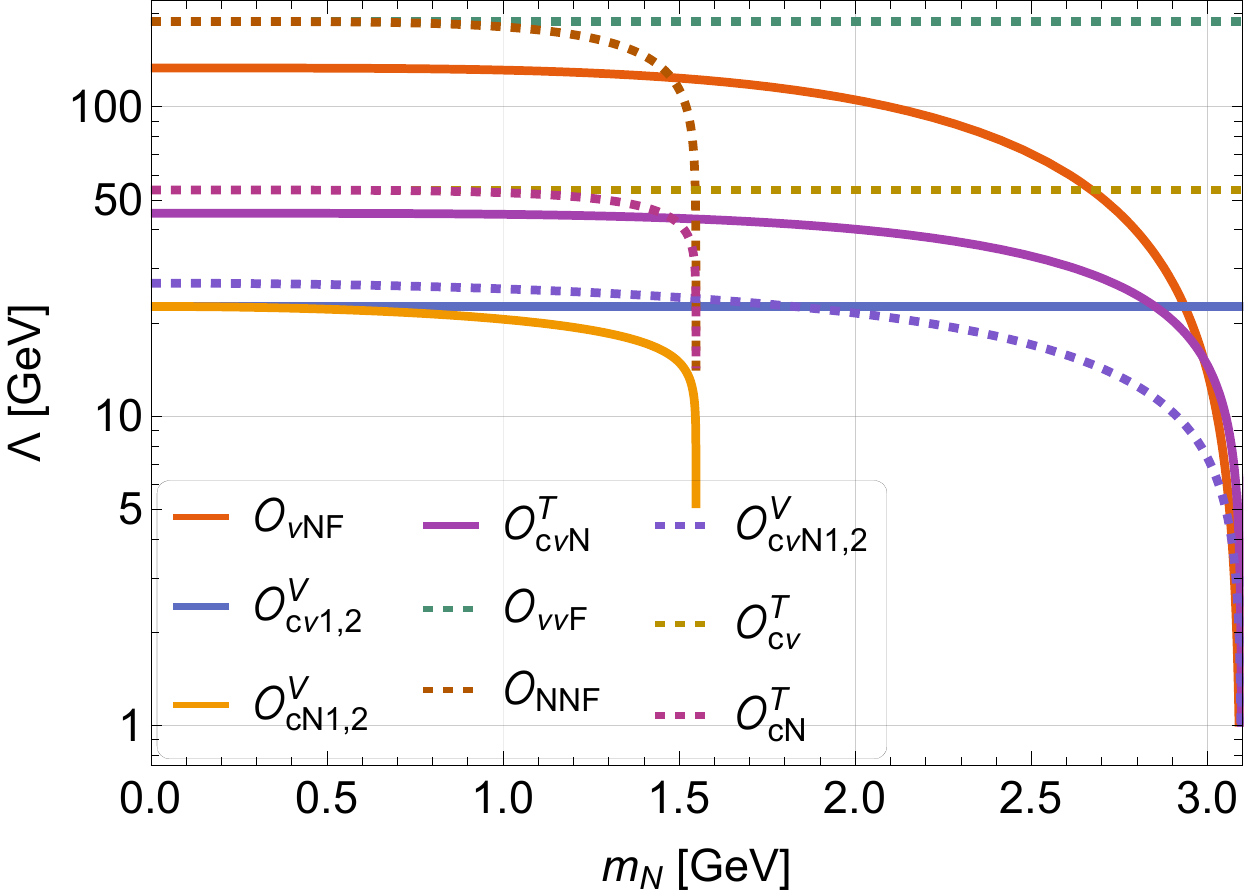}
\includegraphics[width=7.5cm]{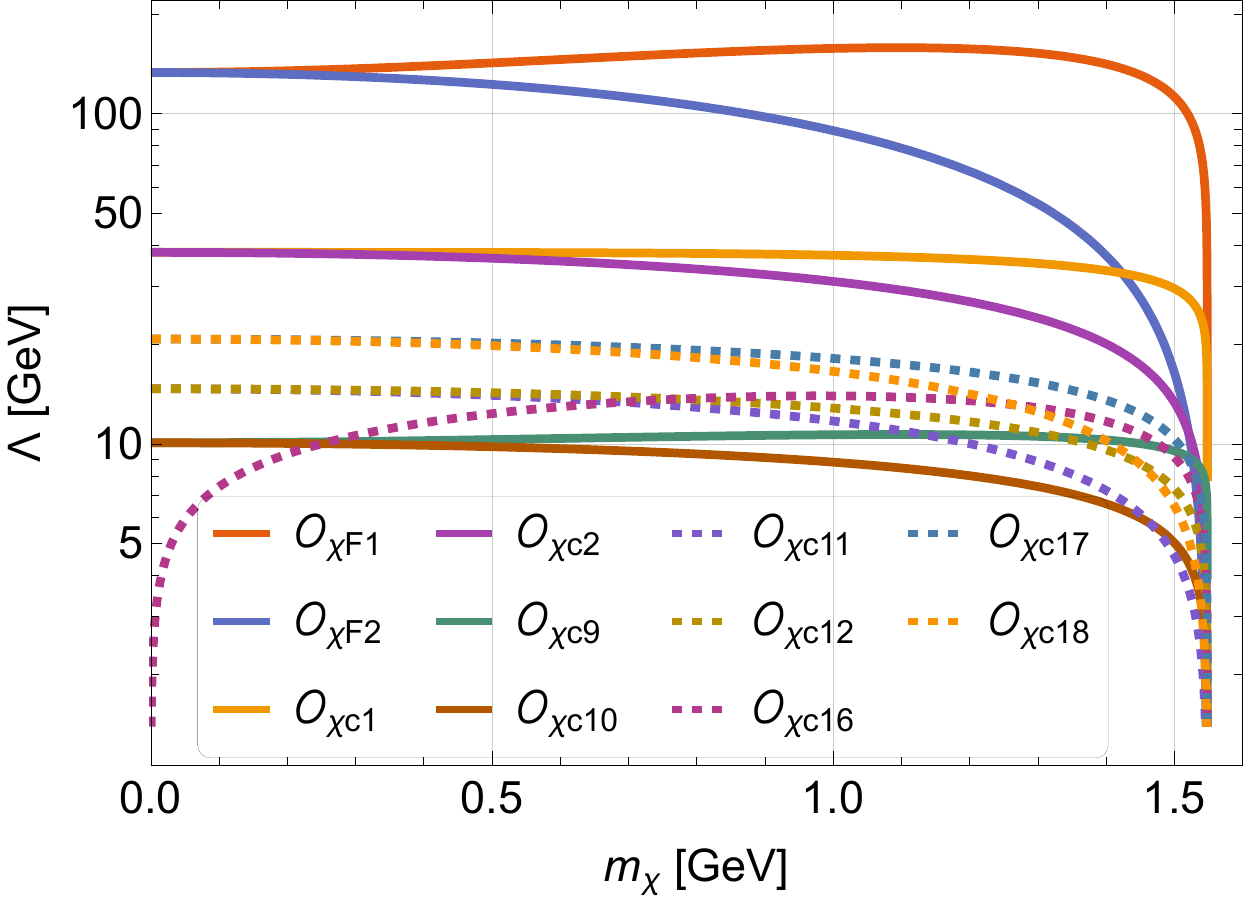}
\caption{Lower limits on the energy scales as a function of the mass of sterile neutrino (left) and DM particle (right) from $J/\psi\to {\rm invisible}$.
}
\label{fig:invlimit}
\end{figure}

In Fig.~\ref{fig:invlimit}, we show the lower constraints on the effective scale as a function of the mass of sterile neutrino (DM particle) in the LNEFT (DMEFT) from $J/\psi\to {\rm invisible}$. Here we have taken into account the current experimental constraint $\mathcal{B}(J/\psi\to {\rm invisible})<7\times 10^{-4}$ and assumed one operator dominant each time to obtain the result. The most stringent bound on the energy scale is above 100 GeV for the dipole operators. Table~\ref{tab:scale_constraints} summarizes the energy bound on individual operator from $J/\psi\to (\gamma+){\rm invisible}$ decays, assuming massless sterile neutrino or DM particle. One can see that the two decay processes provide complementary constraints on the effective operators. The constraint on $\Lambda$ from $J/\psi\to \gamma+{\rm invisible}$ is relatively weak and thus the observation may imply a light degree of freedom for C-even mediator and DM particle in the UV completions.

The dipole operators are also constrained by other meson decays and the CE$\nu$NS process and have been studied by some of us~\cite{Li:2020lba}. In particular the energy scales associated with the dipole operator Wilson coefficients $C_{\nu NF, \nu\nu F}^{\alpha\beta}$ with $\alpha\neq \tau$ are constrained by CE$\nu$NS to be larger than $1900$ TeV and $3700$ TeV, respectively. Invisible $J/\psi$ decays place the stronger constraints on the Wilson coefficients $C_{\nu NF,\nu\nu F}^{\tau\beta}$ and $C_{NNF}^{\alpha\beta}$ for $\alpha,\beta=e,\mu,\tau$ than invisible decays of $\omega$ and $\phi$ vector mesons which have been studied in Ref.~\cite{Li:2020lba}. Constraints on 4-fermion operators are not directly comparable due to the different quarks in the operator and here the searches for invisible $J/\psi$ decays and $J/\psi\to \gamma+$ invisible are complementary.

\begin{table}
\centering
\resizebox{\linewidth}{!}{
\renewcommand{\arraystretch}{1.2}
\begin{tabular}{| c | c | c | c | c | c | }
\hline
LNEFT WC& \multicolumn{2}{|c|}{$\Lambda_{\rm LNEFT}=\left|C_i\right|^{1\over 4-d}\;[\rm GeV]$} &
DMEFT WC &  \multicolumn{2}{|c|}{$\Lambda_{\rm DMEFT}=\left|C_i\right|^{1\over 4-d}\;[\rm GeV]$}
\\\hline
 &  $J/\psi\to \rm invisible$  &  $J/\psi\to \rm\gamma+invisible$  &
 &  $J/\psi\to \rm invisible$  &  $J/\psi\to \rm\gamma+invisible$
\\\hline
$C_{\nu N F}$  &  133.2   & - &  $C_{\chi F1,2}$ & 133.2  & -
\\\hline
$C_{c\nu1,2}^{V}\;,\;C_{cN1,2}^{V}$ & 22.6 &12.8   &   $C_{\chi c1,2}$   & 38.0  & -
\\\hline
$C_{c\nu N1,2}^{S}$ & -  &8.8    &  $C_{\chi c3,4}$  & -  &11.6
\\\hline
$C_{c\nu N}^{T}$ &  45.2   & - & $C_{\chi c5,6}$  & - &  4.0
\\\hline
$C_{\nu\nu F}\;,\;C_{NN F}$ & 188.4  & - & $C_{\chi c7,8}$ & - & 4.4
\\\hline
$C_{c\nu N1,2}^{V}$ &  26.9  &12.8   & $C_{\chi c9,10}$   &  10.1  & -
\\\hline
$C_{c\nu 1,2}^{S}\;,\;C_{cN 1,2}^{S}$ & - &12.4    &  $C_{\chi c11,12}$   & 14.7  & -
\\\hline
$C_{c\nu}^{T}\;,\;C_{cN}^{T}$ & 53.8    & - &  $C_{\chi c13,14}$  & - &4.7
\\\hline
& & & $C_{\chi c15,16}$   & -  & -
\\\hline
& & & $C_{\chi c17,18}$  & 20.8 & -
\\\hline
\end{tabular}
}
\caption{Constraints on the effective scales defined in terms of the Wilson coefficients as $\Lambda\equiv C_{i}^{1\over d-4}$ in the massless limit. For the operators with active neutrinos, the limits are obtained under the assumption of one single-flavor dominance for the flavor dependent couplings. Thus the limits for the WCs involving $\nu$ are applicable for any flavor of active neutrinos, with the flavor indices omitted for simplicity.
}
\label{tab:scale_constraints}
\end{table}

\subsection{The DM direct detection in DMEFT}
\label{sec:DMDD}

The DMEFT operators also determine the cross section of DM scattering off a nucleus in the direct detection experiments.
The lower bound on the scale of the effective operator
obtained above can be converted to the upper limit on the DM-nucleon scattering cross section. Among the operators relevant for the radiative decay of $J/\psi$, only the scalar operator $\mathcal{O}_{\chi q5}$ leads to spin-independent (SI) DM-nucleon scattering cross section which is meanwhile not suppressed by momentum transfer. The axial vector operator $\mathcal{O}_{\chi q4}$ contributes to the spin-dependent (SD) scattering cross section. For the operators relevant for the purely invisible decay of $J/\psi$, the vector operator $\mathcal{O}_{\chi q1}$ typically gives non-momentum-suppressed SI scattering cross section. We next evaluate the non-suppressed DM-nucleon scattering cross sections converted from $J/\psi$ data. Other operators such as $\mathcal{O}_{\chi q2}, \mathcal{O}_{\chi q3}$ at the nucleon level are decomposed into the non-relativistic operators depending on momentum transfer and thus lead to momentum-suppressed scattering cross section~\cite{Fitzpatrick:2012ix,DelNobile:2013sia,Brod:2017bsw}. We will not consider them below.

The form factors of nucleon $\mathbb{N}$ are defined as~\cite{DelNobile:2013sia,Bishara:2017pfq}
\begin{eqnarray}
\langle \mathbb{N}| \bar{q}\gamma_\mu q|\mathbb{N}\rangle&=&c_q^\mathbb{N} \bar{\mathbb{N}}\gamma_\mu\mathbb{N}\;, \quad q=u,d\;,\\
\langle \mathbb{N}| m_q \bar{q}q|\mathbb{N}\rangle&=&m_\mathbb{N} f_q^\mathbb{N} \bar{\mathbb{N}}\mathbb{N}\;, \quad q=u,d,s\;,\\
\langle \mathbb{N}| m_Q \bar{Q}Q|\mathbb{N}\rangle&=&\langle \mathbb{N}| {-\alpha_s\over 12\pi} G^a_{\mu\nu}G^{a\mu\nu}|\mathbb{N}\rangle = {2\over 27} m_\mathbb{N} f_G^\mathbb{N} \bar{\mathbb{N}}\mathbb{N}\;, \quad Q=c,b,t\;,
\end{eqnarray}
for the scalar SI interactions with $m_\mathbb{N}$ being the nucleon mass and $c_u^p=c_d^n=2, c_d^p=c_u^n=1$. Those for SD interactions are
\begin{eqnarray}
\langle \mathbb{N}| \bar{q}\gamma_\mu\gamma_5 q|\mathbb{N}\rangle&=& \Delta_q^\mathbb{N} \bar{\mathbb{N}}\gamma_\mu \gamma_5 \mathbb{N}\;, \quad q=u,d,s\;.
\end{eqnarray}
The elastic SI scattering cross sections from $\mathcal{O}_{\chi q1}$ and $\mathcal{O}_{\chi q5}$ are given by~\cite{Fernandez:2015klv}
\begin{eqnarray}
\sigma_{\chi q1}^{\rm SI}&=& {9\mu_{\chi \mathbb{N}}^2\over \pi} |C_{\chi q1}|^2\;,\\
\sigma_{\chi q5}^{\rm SI}&=& {\mu_{\chi \mathbb{N}}^2\over \pi} |C_{\chi q5}|^2 m_\mathbb{N}^2 \Big(\sum_{q=u,d,s}f_q^\mathbb{N} + {2\over 27}\sum_{q=c,b,t}f_G^\mathbb{N} \Big)^2\;,
\end{eqnarray}
where $\mu_{\chi \mathbb{N}}=m_\chi m_\mathbb{N}/(m_\chi + m_\mathbb{N})$ is the DM-nucleon reduced mass. The SD cross section is
\begin{eqnarray}
\sigma_{\chi q4}^{\rm SD}&=& {3\mu_{\chi \mathbb{N}}^2\over \pi} |C_{\chi q4}|^2 \Big(\sum_{q=u,d,s}\Delta_q^\mathbb{N} \Big)^2\;.
\end{eqnarray}
In the above cross section formulas, we assumed the universal WCs to the SM quarks. By contrast, assuming only charm quark coupling to the DM prevents the contributions in SI cross section from light quarks and the SD cross section. In this case we only have non-vanishing SI scattering cross section from $\mathcal{O}_{\chi c5}$. Next we consider both of these two assumptions and evaluate the limit on the DM-nucleon scattering cross section.

\begin{figure}[htb!]
	\includegraphics[width=0.49\linewidth]{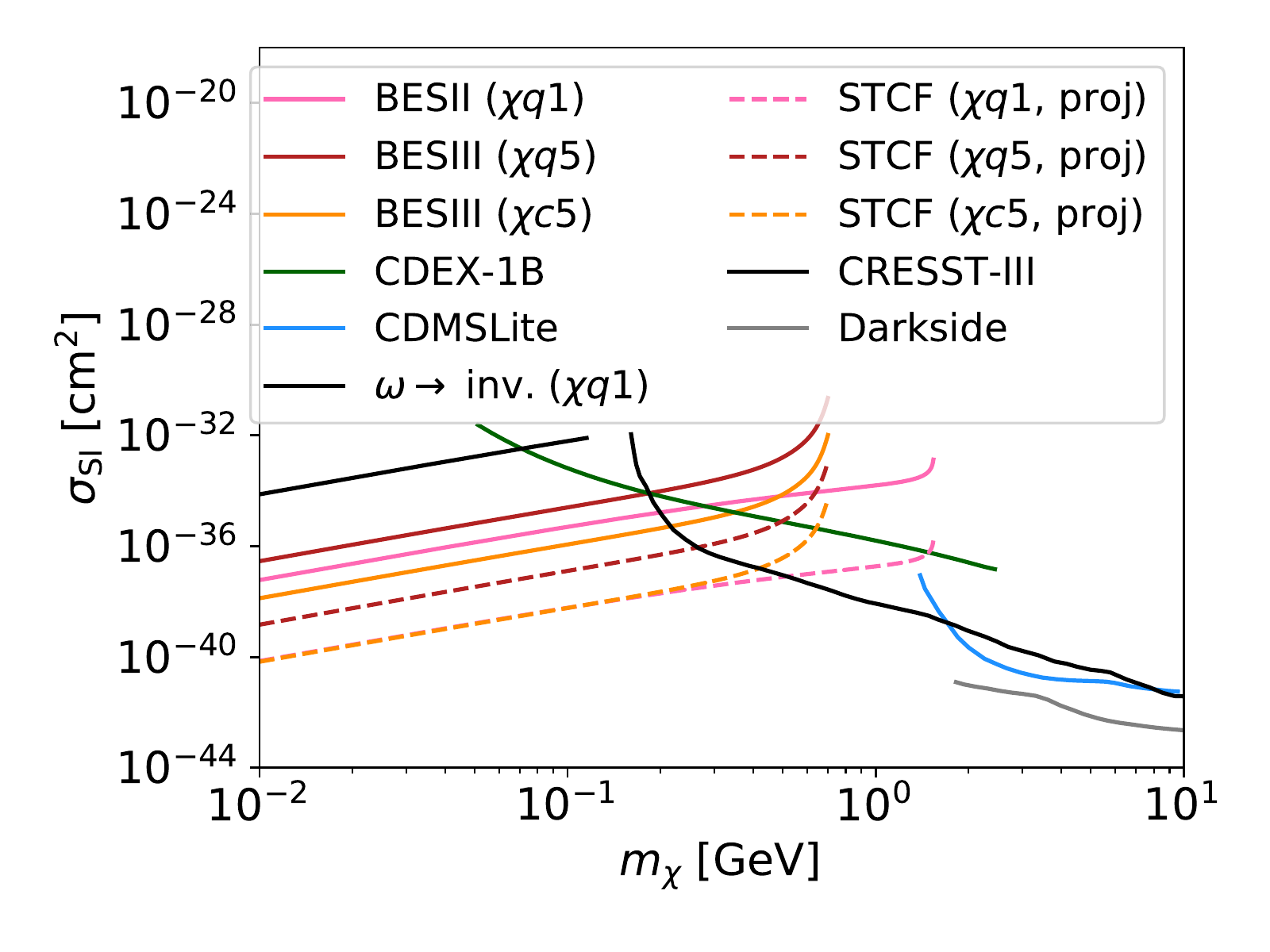}
	\includegraphics[width=0.49\linewidth]{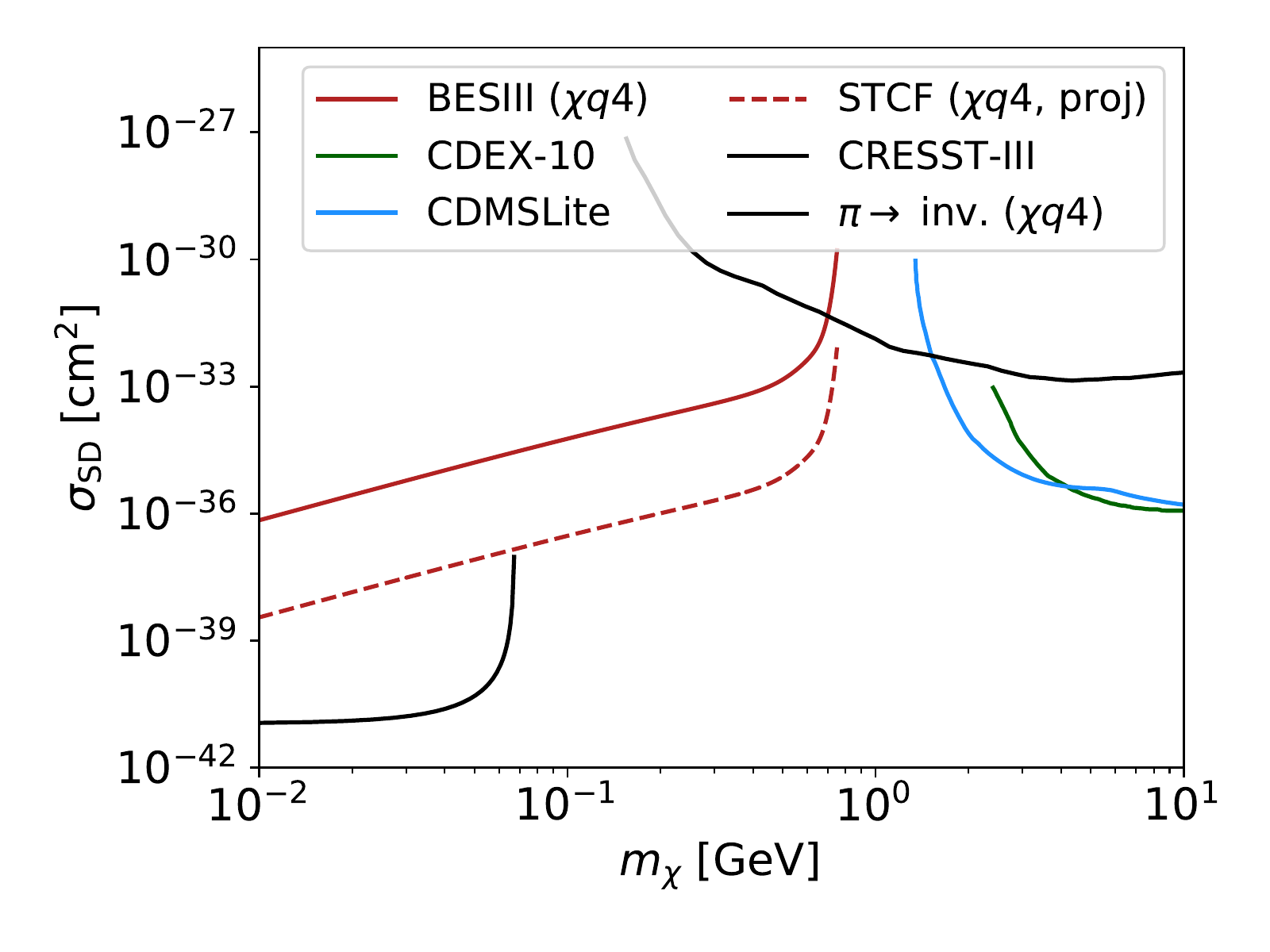}
	\caption{The SI (left) and SD (right) DM-nucleon scattering cross sections converted from $J/\psi$ data (solid) and projection for STCF (dashed). Left: We assume universal quark coupling for operators $\mathcal{O}_{\chi q1}$ (pink) and $\mathcal{O}_{\chi q5}$ (red) as well as only non-vanishing charm quark coupling for operator $\mathcal{O}_{\chi c5}$ (orange). The SI limits from DM direct detection are also shown, including CDEX-1B (green)~\cite{Liu:2019kzq}, CDMSLite (blue)~\cite{Agnese:2017jvy}, CRESST-III (black)~\cite{Abdelhameed:2019hmk} and Darkside (gray)~\cite{Agnes:2018ves}. Invisible $\omega(782)$ decay (black) provides a constraint for  $\mathcal{O}_{\chi q1}$ assuming universal quark couplings.
		Right: We assume universal quark coupling for operator $\mathcal{O}_{\chi q4}$ (red). The SD limits from DM direct detection include CDEX-10 (green)~\cite{Jiang:2018pic}, CDMSLite (blue)~\cite{Agnese:2017jvy} and CRESST-III (black)~\cite{Abdelhameed:2019hmk}. Invisible neutral pion decay (black) places a constraint for operator $\mathcal{O}_{\chi q4}$ assuming universal quark couplings.
	}
	\label{fig:DMDD1}
\end{figure}

In Fig.~\ref{fig:DMDD1} we show the upper limits on the SI and SD DM-nucleon scattering cross sections from the $J/\psi$ constraints obtained above.
One can see that the invisible decays of $J/\psi$ are sensitive to the light DM mass range which cannot be probed yet by DM direct detection experiments. For $m_\chi\ll m_\mathbb{N}$ the SI DM-nucleon scattering cross section scales like $\sigma^{\rm SI}_{\chi q 1}\propto m_\chi^2 |C_{\chi q 1}|^2$ or $\sigma^{\rm SI}_{\chi q 5}\propto m_\chi^2 m_\mathbb{N}^2 |C_{\chi q 5}|^2$ and thus the inferred constraint on the SI DM-nucleon scattering cross section is becoming more stringent for smaller DM masses $m_\chi$.
For the case of only charm quark coupling and $m_\chi=10^{-2}$ GeV for instance, the BESIII limit of cross section $\sigma^{\rm SI}_{\chi c 5}$ becomes $1.2\times 10^{-38}~{\rm cm}^2$ and the STCF with $3.4\times 10^{12}$ samples of $J/\psi$~\cite{STCF} can reach a sensitivity down to $6.3\times 10^{-41}~{\rm cm}^2$ after one year of running. Under the assumption of universal quark coupling, the SI cross section limit for $\mathcal{O}_{\chi q1}$ from $J/\psi\to {\rm invisible}$ is stronger by one order of magnitude than that for $\mathcal{O}_{\chi q5}$ from $J/\psi\to \gamma+{\rm invisible}$.
For the SD cross section with universal quark coupling and $m_\chi=10^{-2}$ GeV, the BESIII limit reaches $6.5\times 10^{-37}~{\rm cm}^2$ and the STCF projection is $3.3\times 10^{-39}~{\rm cm}^2$.

Under the assumption of universal quark coupling, there exist additional constraints on the DMEFT coefficients from the invisible decay of light mesons. Recently, NA62 placed a strong constraint on invisible pion decay with $\mathcal{B}(\pi^0\to\mathrm{inv.})<4.4\times 10^{-9}$~\cite{NA62:2020pwi}, about two orders of magnitude more stringent than the previous bound. Among those DMEFT operators, only the operator ${\cal O}_{\chi q4}$ can contribute to both the pseudoscalar $\pi^{0}$ invisible decay and the non-momentum-suppressed SD DM-nucleon scattering. The invisible decay $\pi^{0}\to \chi\bar\chi$ can only take place if the DM matter $m_{\chi} \leq m_{\pi}/2\approx67.5$~MeV. The
branching ratio due to ${\cal O}_{q\chi4}$ is given by
\begin{eqnarray}
	\mathcal{B}(\pi^{0}\to\chi\bar\chi)
	={\tau_{\pi} m_{\pi} \over 8 \pi }m_{\chi}^{2}f_{\pi}^{2}|C_{\chi u4}+C_{\chi d4}|^{2}
	\left(1-4\frac{m_{\chi}^{2}}{m_{\pi}^{2}} \right)^{2}\; ,
\end{eqnarray}
where the pion decay constant is $f_{\pi}=130.2$~MeV, and $\tau_{\pi}=(8.43\pm0.13)\times 10^{-17}$ s is the $\pi^{0}$ lifetime.
Hence for DM masses $m_\chi<m_\pi/2$, the invisible pion decay imposes the most stringent constraint and excludes the SD DM-nucleon scattering cross section above $1.1\times 10^{-41}~\mathrm{cm}^2$ for $m_\chi=10^{-2}$ GeV.
The invisible decay of vector mesons such as $\omega(782)$ also constrains the operator ${\cal O}_{\chi q1}$ with $\mathcal{B}(\omega(782)\to\mathrm{inv.})<7.0\times 10^{-5}$~\cite{Zyla:2020zbs}. The branching ratio due to $\mathcal{O}_{q\chi1}$ is given by
\begin{equation}
	\mathcal{B}(\omega(782)\to \chi\bar\chi) = \frac{ m_\omega^3 f_\omega^2}{24\pi\Gamma_\omega}  \left|C_{\chi u1} + C_{\chi d1}\right|^2 \left(1+2\frac{m_\chi^2}{m_\omega^2}\right) \left(1-4\frac{m_\chi^2}{m_\omega^2}\right)^{1/2}\;,
\end{equation}
where the $\omega(782)$ decay constant is $f_\omega=187$ MeV, and $\Gamma_\omega=(8.68\pm0.13)$ MeV is the width of $\omega(782)$. The constraint on the SI scattering cross section from invisible $\omega(782)$ decay for $\mathcal{O}_{\chi q1}$ is several orders of magnitude weaker than the one from invisible $J/\psi$ decay.

\section{Conclusions}
\label{sec:Con}

The heavy quarkonium experiments can help us to study the possible NP associated with heavy quarks and provide complementary constraints on the NP scale where the high-energy colliders lose sensitivity.
Inspired by the searches for $J/\psi$ decays into invisible particles, we investigate the implication for light sterile neutrino and sub-GeV dark matter in effective field theories.

We make use of the low-energy EFTs for general neutrino operators up to dim-6 and the Dirac fermion DM operators up to dim-7.
For $J/\psi\to \gamma+{\rm invisible}$ decay, we perform the likelihood fits for the individual LNEFT and DMEFT operators with distinct Lorentz structures and photon spectra.
The limits on the decay branching fractions are obtained for different neutrino or DM scenarios and then converted to the lower bounds on the new energy scales.
The most stringent bound on the energy scale in LNEFT comes from the lepton-number-conserving operators $\mathcal{O}_{c\nu 1,2}^V$ and turns out to be 12.8 GeV.
For DMEFT, the most stringent bound on the energy scale is 11.6 GeV for the axialvector operators $\mathcal{O}_{c\chi 3,4}$.
The purely invisible decay $J/\psi\to {\rm invisible}$ provides complementary constraints on the effective operators. The most stringent bound on the energy scale is above 100 GeV for the dipole operators.

Finally, we evaluate the limit on the DM-nucleon scattering cross section converted from $J/\psi$ data. The invisible decay of $J/\psi$ is sensitive to the light DM mass range where the DM direct detection experiments cannot probe yet.
For the case of only charm quark coupling and $m_\chi=10^{-2}$ GeV for instance, the BESIII limit of cross section $\sigma^{\rm SI}_{\chi c 5}$ becomes $1.2\times 10^{-38}~{\rm cm}^2$ and the STCF with $3.4\times 10^{12}$ samples of $J/\psi$~\cite{STCF} can reach a sensitivity down to $6.3\times 10^{-41}~{\rm cm}^2$ after one year of running. Under the assumption of universal quark coupling, the SI cross section limit for $\mathcal{O}_{\chi q1}$ from $J/\psi\to {\rm invisible}$ is stronger by one order of magnitude than that for $\mathcal{O}_{\chi q5}$ from $J/\psi\to \gamma+{\rm invisible}$.
For the SD cross section under the assumption of universal quark coupling and $m_\chi=10^{-2}$ GeV, the BESIII limit reaches $6.5\times 10^{-37}~{\rm cm}^2$ and the STCF projection is $3.3\times 10^{-39}~{\rm cm}^2$.

\acknowledgments
TL would like to thank Xiao-Dong Shi and Ming-Gang Zhao for very useful discussions. MS acknowledges useful discussions with Yi Cai.
TL is supported by the National Natural Science Foundation of China (Grant No. 11975129, 12035008) and ``the Fundamental Research Funds for the Central Universities'', Nankai University (Grants No. 63196013). XDM is supported by Shanghai Pujiang Program (20PJ1407800), and National Natural Science Foundation of China (No. 12090064). MS acknowledges support by the Australian Research Council via the Discovery Project DP200101470.

\appendix

\section{The proof of redundant DM operators}
\label{sec:DMredundant}

We prove that the four DM operators $\hat{\cal O}_{\chi q15},\hat{\cal O}_{\chi q16},\hat{\cal O}_{\chi q17},\hat{\cal O}_{\chi q18}$ in Eq.~\eqref{eq: red_opes} are actually redundant operators which can be shifted into the other operators by
the Dirac gamma matrix identity (GI) and the equation of motion (EoM) of DM fields. They all have a derivative acting on the DM current such as
\begin{eqnarray}\nonumber
\partial_\mu (\overline{\chi}\sigma^{\mu\nu}\chi)
&=& (\overline{\chi}\overleftarrow{\partial_\mu}\sigma^{\mu\nu}\chi)+(\overline{\chi}\sigma^{\mu\nu}\partial_\mu\chi)
\\\nonumber
&\overset{\rm GI}{=}&(\overline{\chi}i\overleftrightarrow{\partial}^\nu\chi)+(\overline{\chi}i\overleftarrow{\slashed{\partial}}\gamma^\nu\chi)-(\overline{\chi}\gamma^\nu i\slashed{\partial}\chi)
\\
&\overset{\rm EoM}{=}&(\overline{\chi}i\overleftrightarrow{\partial}^\nu\chi)-2m_\chi (\overline{\chi}\gamma^\nu\chi)\;,
\\\nonumber
\partial_\mu (\overline{\chi}\sigma^{\mu\nu}\gamma_5\chi)
&=& (\overline{\chi}\overleftarrow{\partial_\mu}\sigma^{\mu\nu}\gamma_5\chi)+(\overline{\chi}\gamma_5\sigma^{\mu\nu}\partial_\mu\chi)
\\\nonumber
&\overset{\rm GI}{=}&(\overline{\chi}i\overleftrightarrow{\partial}^\nu\gamma_5\chi)+(\overline{\chi}i\overleftarrow{\slashed{\partial}}\gamma^\nu\gamma_5\chi)-(\overline{\chi}\gamma_5\gamma^\nu i\slashed{\partial}\chi)
\\
&\overset{\rm EoM}{=}&(\overline{\chi}i\overleftrightarrow{\partial}^\nu\gamma_5\chi)\;.
\end{eqnarray}
Then, it is straightforward to obtain
\begin{align}
\hat{\cal O}_{\chi q15}&={\cal O}_{\chi q11}-2m_\chi {\cal O}_{\chi q1}\;,
&\hat{\cal O}_{\chi q16}&={\cal O}_{\chi q12}\;,
\\
\hat{\cal O}_{\chi q17}&={\cal O}_{\chi q13}-2m_\chi {\cal O}_{\chi q3}\;,
&\hat{\cal O}_{\chi q18}&={\cal O}_{\chi q14}\;.
\end{align}
The above finishes the proof of the redundancy.

The missed four independent operators are
\begin{align}\nonumber
{\cal O}_{\chi q15}&=(\overline{\chi}\gamma^{[\mu}i\overleftrightarrow{\partial}^{\nu]}\chi)(\overline{q}\sigma_{\mu\nu} q)\;,&
{\cal O}_{\chi q16}&=(\overline{\chi}\gamma^{[\mu}i\overleftrightarrow{\partial}^{\nu]}\gamma_5\chi)(\overline{q}\sigma_{\mu\nu} q)\;,
\\
{\cal O}_{\chi q17}&=(\overline{\chi}\gamma^{[\mu}i\overleftrightarrow{\partial}^{\nu]} \chi)(\overline{q}\sigma_{\mu\nu}i \gamma_5 q)\;,&
{\cal O}_{\chi q18}&=(\overline{\chi}\gamma^{[\mu}i\overleftrightarrow{\partial}^{\nu]}\gamma_5\chi)(\overline{q}\sigma_{\mu\nu}i\gamma_5 q)\;,
\end{align}
where $\gamma^{[\mu}\overleftrightarrow{\partial}^{\nu]}$ indicates the anti-symmetrization of the two Lorentz indices $\mu$ and $\nu$.
This parametrization is easily formulated in the framework of the chiral effective field with the DM currents together with the relevant Wilson coefficients treated as the tensor external sources~\cite{Cata:2007ns}.
Or equivalently, we can arrange the derivatives acting on the quark field and parameterize the missed four operators as
\begin{align}\nonumber
&\tilde{\cal O}_{\chi q15}=(\overline{\chi}\gamma_\mu\chi)(\overline{q}i\overleftrightarrow{D}^\mu q)\;,
&&\tilde{\cal O}_{\chi q16}=(\overline{\chi}\gamma_\mu \gamma_5\chi)(\overline{q}i\overleftrightarrow{D}^\mu q)\;,
\\
&\tilde{\cal O}_{\chi q17}=(\overline{\chi}\gamma_\mu\chi)(\overline{q}i\overleftrightarrow{D}^\mu i\gamma_5 q)\;,
&&\tilde{\cal O}_{\chi q18}=(\overline{\chi}\gamma_\mu \gamma_5\chi)(\overline{q}i\overleftrightarrow{D}^\mu i\gamma_5 q)\;.
\end{align}
The latter parametrization has a similar structure as the operators ${\cal O}_{\chi q11,12,13,14}$. The equivalence can be easily established by exploiting the integration by parts relations, the EoMs of DM and quark fields and the above Dirac gamma matrix identity together with $\gamma^{\alpha} \gamma^{\beta} \gamma^{\rho}=g^{\alpha \beta} \gamma^{\rho}+g^{\beta \rho} \gamma^{\alpha}-g^{\alpha \rho} \gamma^{\beta}+i \varepsilon^{\alpha \beta \rho \nu} \gamma_{\nu} \gamma_{5}$. The relationship is as follows
\begin{align}\nonumber
{\cal O}_{\chi q15}&=2\tilde{\cal O}_{\chi q18}\;,
&{\cal O}_{\chi q16}&=2{m_\chi\over m_q}{\cal O}_{\chi q10}+2\tilde{\cal O}_{\chi q17}\;,
\\
{\cal O}_{\chi q17}&=4m_q{\cal O}_{\chi q2}+2\tilde{\cal O}_{\chi q16}\;,
&{\cal O}_{\chi q18}&=4m_q{\cal O}_{\chi q1}-2{m_\chi\over m_q}{\cal O}_{\chi q9}-2\tilde{\cal O}_{\chi q15}\;.
\end{align}
We take ${\cal O}_{\chi q15,16,17,18}$ as our basis operators. Unlike the original operators in Ref.~\cite{Brod:2017bsw}, now these new operators have non-vanishing 1-loop QCD renormalization and the anomalous dimension is the same as the dim-6 tensor operators.

\section{\boldmath The matrix elements of $J/\psi$ decay in LNEFT}
\label{sec:ME-LNEFT}

The relevant LNEFT interactions up to dim-6, for the process $J/\psi(P)\to \gamma(k)+{\rm inv}_{1/\alpha}(k_1)+{\rm inv}_{2/\beta}(k_2)$, lead to the following LNC amplitudes
\begin{eqnarray}
 \mathcal{M}(J/\psi\to\gamma \nu_\alpha\bar{\nu}_\beta)&=&m_J{\cal N}(q^2)i\epsilon^{\mu\nu\rho\sigma}k_\nu\epsilon_{J,\rho}\epsilon_{\gamma,\sigma}^*\left(C_{q\nu 1}^{V,\alpha\beta}-C_{q\nu 2}^{V,\alpha\beta}\right) \overline{u_\nu} \gamma^\mu P_L v_{\bar \nu} \;,
\\
\mathcal{M}(J/\psi\to\gamma N_\alpha\bar{N}_\beta)&=&m_J{\cal N}(q^2)i\epsilon^{\mu\nu\rho\sigma}k_\nu\epsilon_{J,\rho}\epsilon_{\gamma,\sigma}^*  \left(C_{qN 1}^{V,\alpha\beta} -C_{qN 2}^{V,\alpha\beta}\right) \overline{u_N} \gamma^\mu P_R v_{\bar N} \;,
\\\nonumber
\mathcal{M}(J/\psi\to\gamma \nu_\alpha\bar{N}_\beta)&=&{\cal N}(q^2)\left[\left((P\cdot\epsilon_\gamma^*)(k\cdot \epsilon_J)-(P\cdot k)(\epsilon_J\cdot \epsilon_\gamma^*) \right)
\left( C_{q\nu N1}^{S,\alpha\beta}+ C_{q\nu N2}^{S,\alpha\beta}\right) \right.
\\
&& \left. +i \epsilon^{\mu\nu\rho\sigma}P_\mu k_\nu\epsilon_{J,\rho}\epsilon_{\gamma,\sigma}^*
 \left( C_{q\nu N1}^{S,\alpha\beta}-C_{q\nu N2}^{S,\alpha\beta}\right) \right]
 \overline{u_\nu} P_R v_{\bar N} \;,
\\\nonumber
  \mathcal{M}(J/\psi\to\gamma\bar{\nu}_\alpha N_\beta)&=&{\cal N}(q^2)\left[\left((P\cdot\epsilon_\gamma^*)(k\cdot \epsilon_J)-(P\cdot k)(\epsilon_J\cdot \epsilon_\gamma^*) \right)
 \left(C_{q\nu N1}^{S,\alpha\beta\ast}+ C_{q\nu N2}^{S,\alpha\beta\ast}\right) \right.
\\
&& \left. -i \epsilon^{\mu\nu\rho\sigma}P_\mu k_\nu\epsilon_{J,\rho}\epsilon_{\gamma,\sigma}^*
 \left(C_{q\nu N1}^{S,\alpha\beta\ast}-C_{q\nu N2}^{S,\alpha\beta\ast}\right) \right]
  \overline{u_N} P_L v_{\bar \nu}\;,
\end{eqnarray}
where $P_{L(R)}=P_{-(+)}$.
The LNV amplitudes with $\Delta L=-2$ are
\begin{eqnarray}\nonumber
\mathcal{M}(J/\psi\to\gamma\bar{\nu}_\alpha\bar{\nu}_\beta)&=& 2{\cal N}(q^2)\left[\left((P\cdot\epsilon_\gamma^*)(k\cdot \epsilon_J)-(P\cdot k)(\epsilon_J\cdot \epsilon_\gamma^*) \right)
\left(C_{q\nu 1}^{S,\alpha\beta} +C_{q\nu 2}^{S,\alpha\beta} \right) \right.
\\
&& \left. -i \epsilon^{\mu\nu\rho\sigma}P_\mu k_\nu\epsilon_{J,\rho}\epsilon_{\gamma,\sigma}^*
\left(C_{q\nu 1}^{S,\alpha\beta} -C_{q\nu 2}^{S,\alpha\beta} \right) \right]
\overline{v^C_{\bar \nu}} P_L v_{\bar \nu^\prime}\;,
\\\nonumber
\mathcal{M}(J/\psi\to\gamma\bar{N}_\alpha\bar{N}_\beta)&=& 2{\cal N}(q^2)\left[\left((P\cdot\epsilon_\gamma^*)(k\cdot \epsilon_J)-(P\cdot k)(\epsilon_J\cdot \epsilon_\gamma^*) \right)
\left(C_{qN 1}^{S,\alpha\beta} + C_{qN 2}^{S,\alpha\beta}\right) \right.
\\
&& \left. -i \epsilon^{\mu\nu\rho\sigma}P_\mu k_\nu\epsilon_{J,\rho}\epsilon_{\gamma,\sigma}^*
\left(C_{qN 1}^{S,\alpha\beta} - C_{qN 2}^{S,\alpha\beta}\right) \right]
 \overline{v^C_{\bar N}}  P_R v_{\bar N^\prime}\;,
\\
\mathcal{M}(J/\psi\to\gamma\bar{\nu}_\alpha\bar{N}_\beta)&=& -m_J{\cal N}(q^2) i\epsilon^{\mu\nu\rho\sigma}k_\nu\epsilon_{J,\rho}\epsilon_{\gamma,\sigma}^* \left(C_{q\nu N1}^{V,\alpha\beta}-C_{q\nu N2}^{V,\alpha\beta}\right) \overline{v^C_{\bar N}} \gamma^\mu P_Lv_{\bar \nu} \;,
\end{eqnarray}
where $v_{\bar \nu} (v_{\bar N})$ and $v_{\bar \nu^\prime} (v_{\bar N^\prime})$ are the spinors of anti-neutrinos $\bar \nu_\alpha (\bar N_\alpha)$ and $\bar \nu_\beta (\bar N_\beta)$, respectively. The amplitudes with $\Delta L=2$ are
\begin{eqnarray}\nonumber
\mathcal{M}(J/\psi\to\gamma\nu_\alpha \nu_\beta)&=& 2{\cal N}(q^2)\left[\left((P\cdot\epsilon_\gamma^*)(k\cdot \epsilon_J)-(P\cdot k)(\epsilon_J\cdot \epsilon_\gamma^*) \right)
\left(C_{q\nu 1}^{S,\alpha\beta\ast} +C_{q\nu 2}^{S,\alpha\beta\ast}\right) \right.
\\
&& \left. +i \epsilon^{\mu\nu\rho\sigma}P_\mu k_\nu\epsilon_{J,\rho}\epsilon_{\gamma,\sigma}^*
\left(C_{q\nu 1}^{S,\alpha\beta\ast} -C_{q\nu 2}^{S,\alpha\beta\ast}\right) \right]
\overline{u_\nu} P_Ru^C_{\nu^\prime}\;,
\\\nonumber
\mathcal{M}(J/\psi\to\gamma N_\alpha N_\beta)&=& 2{\cal N}(q^2)\left[\left((P\cdot\epsilon_\gamma^*)(k\cdot \epsilon_J)-(P\cdot k)(\epsilon_J\cdot \epsilon_\gamma^*) \right)
\left(C_{qN 1}^{S,\alpha\beta\ast} +C_{qN 2}^{S,\alpha\beta\ast}\right) \right.
\\
&& \left. +i \epsilon^{\mu\nu\rho\sigma}P_\mu k_\nu\epsilon_{J,\rho}\epsilon_{\gamma,\sigma}^*
\left(C_{qN 1}^{S,\alpha\beta\ast} -C_{qN 2}^{S,\alpha\beta\ast}\right) \right]
\overline{u_N} P_L u^C_{N^\prime}\;,
\\
 \mathcal{M}(J/\psi \to \gamma \nu_\alpha N_\beta)&=&-m_J{\cal N}(q^2)i\epsilon^{\mu\nu\rho\sigma}k_\nu\epsilon_{J,\rho}\epsilon_{\gamma,\sigma}^* \left(C_{q\nu N1}^{V,\alpha\beta\ast}- C_{q\nu N2}^{V,\alpha\beta\ast}\right)\overline{u_\nu} \gamma^\mu P_Lu^C_{N} \;,
\end{eqnarray}
where $u_{\bar \nu} (u_{\bar N})$ and $u_{\bar \nu^\prime} (u_{\bar N^\prime})$ are the spinors of neutrinos $\nu_\alpha (N_\alpha)$ and $\nu_\beta (N_\beta)$, respectively.

We then define the kinematic functions
\begin{align}
f(\mu_+)&=1-x_\gamma-\mu_+\;,
\\\nonumber
g(\mu_-)&={1\over 2x_\gamma^2}\left[2(1-x_b+\mu_-)(1-x_c-\mu_-)
+x_\gamma\left( (1-x_b+\mu_-)x_b+(1-x_c-\mu_-)x_c\right)
\right]\;,
\label{fgfunction}
\end{align}
where $\mu_\pm\equiv\mu_b\pm \mu_c=\mu_\alpha\pm \mu_\beta$.
With the above definitions and abbreviations, the spin-averaged squared amplitudes in LNC case become
\begin{eqnarray}\nonumber
\overline{|\mathcal{M}_{\nu_\alpha\bar{\nu}_\beta}|}^2&=&{16e^2Q_c^2|\Psi(0)|^2m_J}
\left|C_{q\nu 1}^{V,\alpha\beta}-C_{q\nu 2}^{V,\alpha\beta}\right|^2
g(\mu_-)\;,
\\\nonumber
\overline{|\mathcal{M}_{N_\alpha\bar{N}_\beta}|}^2&=&{16e^2Q_c^2|\Psi(0)|^2m_J}
 \left|C_{qN 1}^{V,\alpha\beta} -C_{qN 2}^{V,\alpha\beta}\right|^2g(\mu_-)\;,
\\
\overline{|\mathcal{M}_{\nu_\alpha\bar{N}_\beta}|}^2&=&{16e^2Q_c^2|\Psi(0)|^2m_J}\left( \left|C_{q\nu N1}^{S,\alpha\beta}\right|^2+ \left|C_{q\nu N2}^{S,\alpha\beta}\right|^2 \right)f(\mu_+)=\overline{ |\mathcal{M}_{\bar{\nu}_\alpha N_\beta}|}^2\;.
\end{eqnarray}
The LNV cases with $|\Delta L|=2$ are
\begin{eqnarray}\nonumber
\overline{|\mathcal{M}_{\bar{\nu}_\alpha\bar{\nu}_\beta}|}^2
&=&{64e^2Q_c^2|\Psi(0)|^2m_J}\left( \left|C_{q\nu 1}^{S,\alpha\beta} \right|^2+ \left|C_{q\nu 2}^{S,\alpha\beta}\right|^2 \right)f(\mu_+)=\overline{|\mathcal{M}_{\nu_\alpha \nu_\beta}|}^2 \;,
\\\nonumber
\overline{|\mathcal{M}_{\bar{N}_\alpha\bar{N}_\beta}|}^2
&=&{64e^2Q_c^2|\Psi(0)|^2m_J}\left( \left|C_{qN 1}^{S,\alpha\beta} \right|^2+ \left|C_{qN 2}^{S,\alpha\beta}\right|^2 \right)f(\mu_+)=\overline{|\mathcal{M}_{N_\alpha N_\beta}|}^2\;,
\\
\overline{|\mathcal{M}_{\bar{\nu}_\alpha\bar{N}_\beta}|}^2
&=&{16e^2Q_c^2|\Psi(0)|^2m_J}
\left|C_{q\nu N1}^{V,\alpha\beta}-C_{q\nu N2}^{V,\alpha\beta}\right|^2g(\mu_-)=\overline{|\mathcal{M}_{\nu_\alpha N_\beta}|}^2 \;.
\end{eqnarray}
The integration over the kinematic functions results in the distribution functions
\begin{align}\nonumber
F(\mu_+,\mu_-)=\int d x_b f(\mu_+)&={x_\gamma\sqrt{(1-x_\gamma)(1-x_\gamma-2\mu_+)+\mu_-^2}\over 1-x_\gamma}(1-x_\gamma-\mu_+)\;,
\\\nonumber
G(\mu_+,\mu_-)=\int d x_b g(\mu_-)&={x_\gamma\sqrt{(1-x_\gamma)(1-x_\gamma-2\mu_+)+\mu_-^2}\over 6(1-x_\gamma)^3}
\\
&\times\left(2(1-x_\gamma)^2(2-x_\gamma)+(1-x_\gamma^2)\mu_+-(5-x_\gamma)\mu_-^2\right)\;.
\end{align}
For massless neutrinos, they are simplified to be
\begin{align}
&F(0,0)=x_\gamma(1-x_\gamma)={2\over m_J^2}E_\gamma(m_J-2E_\gamma)\;,\;\;
G(0,0)={1\over 3}x_\gamma(2-x_\gamma)={4\over 3m_J^2}E_\gamma(m_J-E_\gamma)\;.
\end{align}

\bibliography{refs}

\begin{thebibliography}{72}%
\makeatletter
\providecommand \@ifxundefined [1]{%
 \@ifx{#1\undefined}
}%
\providecommand \@ifnum [1]{%
 \ifnum #1\expandafter \@firstoftwo
 \else \expandafter \@secondoftwo
 \fi
}%
\providecommand \@ifx [1]{%
 \ifx #1\expandafter \@firstoftwo
 \else \expandafter \@secondoftwo
 \fi
}%
\providecommand \natexlab [1]{#1}%
\providecommand \enquote  [1]{``#1''}%
\providecommand \bibnamefont  [1]{#1}%
\providecommand \bibfnamefont [1]{#1}%
\providecommand \citenamefont [1]{#1}%
\providecommand \href@noop [0]{\@secondoftwo}%
\providecommand \href [0]{\begingroup \@sanitize@url \@href}%
\providecommand \@href[1]{\@@startlink{#1}\@@href}%
\providecommand \@@href[1]{\endgroup#1\@@endlink}%
\providecommand \@sanitize@url [0]{\catcode `\\12\catcode `\$12\catcode
  `\&12\catcode `\#12\catcode `\^12\catcode `\_12\catcode `\%12\relax}%
\providecommand \@@startlink[1]{}%
\providecommand \@@endlink[0]{}%
\providecommand \url  [0]{\begingroup\@sanitize@url \@url }%
\providecommand \@url [1]{\endgroup\@href {#1}{\urlprefix }}%
\providecommand \urlprefix  [0]{URL }%
\providecommand \Eprint [0]{\href }%
\providecommand \doibase [0]{http://dx.doi.org/}%
\providecommand \selectlanguage [0]{\@gobble}%
\providecommand \bibinfo  [0]{\@secondoftwo}%
\providecommand \bibfield  [0]{\@secondoftwo}%
\providecommand \translation [1]{[#1]}%
\providecommand \BibitemOpen [0]{}%
\providecommand \bibitemStop [0]{}%
\providecommand \bibitemNoStop [0]{.\EOS\space}%
\providecommand \EOS [0]{\spacefactor3000\relax}%
\providecommand \BibitemShut  [1]{\csname bibitem#1\endcsname}%
\let\auto@bib@innerbib\@empty
\bibitem [{\citenamefont {Fukuda}\ \emph {et~al.}(1998)\citenamefont {Fukuda}
  \emph {et~al.}}]{Fukuda:1998mi}%
  \BibitemOpen
  \bibfield  {author} {\bibinfo {author} {\bibfnamefont {Y.}~\bibnamefont
  {Fukuda}} \emph {et~al.} (\bibinfo {collaboration} {Super-Kamiokande}),\
  }\href {\doibase 10.1103/PhysRevLett.81.1562} {\bibfield  {journal} {\bibinfo
   {journal} {Phys. Rev. Lett.}\ }\textbf {\bibinfo {volume} {81}},\ \bibinfo
  {pages} {1562} (\bibinfo {year} {1998})},\ \Eprint
  {http://arxiv.org/abs/hep-ex/9807003} {arXiv:hep-ex/9807003} \BibitemShut
  {NoStop}%
\bibitem [{\citenamefont {Aprile}\ \emph {et~al.}(2020)\citenamefont {Aprile}
  \emph {et~al.}}]{Aprile:2020tmw}%
  \BibitemOpen
  \bibfield  {author} {\bibinfo {author} {\bibfnamefont {E.}~\bibnamefont
  {Aprile}} \emph {et~al.} (\bibinfo {collaboration} {XENON}),\ }\href
  {\doibase 10.1103/PhysRevD.102.072004} {\bibfield  {journal} {\bibinfo
  {journal} {Phys. Rev. D}\ }\textbf {\bibinfo {volume} {102}},\ \bibinfo
  {pages} {072004} (\bibinfo {year} {2020})},\ \Eprint
  {http://arxiv.org/abs/2006.09721} {arXiv:2006.09721 [hep-ex]} \BibitemShut
  {NoStop}%
\bibitem [{\citenamefont {Zhou}\ \emph {et~al.}(2020)\citenamefont {Zhou} \emph
  {et~al.}}]{Zhou:2020bvf}%
  \BibitemOpen
  \bibfield  {author} {\bibinfo {author} {\bibfnamefont {X.}~\bibnamefont
  {Zhou}} \emph {et~al.} (\bibinfo {collaboration} {PandaX-II}),\ }\href
  {\doibase 10.1088/0256-307X/38/1/011301} {\  (\bibinfo {year} {2020}),\
  10.1088/0256-307X/38/1/011301},\ \Eprint {http://arxiv.org/abs/2008.06485}
  {arXiv:2008.06485 [hep-ex]} \BibitemShut {NoStop}%
\bibitem [{\citenamefont {Diaz}\ \emph {et~al.}(2020)\citenamefont {Diaz},
  \citenamefont {Arg\"uelles}, \citenamefont {Collin}, \citenamefont {Conrad},\
  and\ \citenamefont {Shaevitz}}]{Diaz:2019fwt}%
  \BibitemOpen
  \bibfield  {author} {\bibinfo {author} {\bibfnamefont {A.}~\bibnamefont
  {Diaz}}, \bibinfo {author} {\bibfnamefont {C.~A.}\ \bibnamefont
  {Arg\"uelles}}, \bibinfo {author} {\bibfnamefont {G.~H.}\ \bibnamefont
  {Collin}}, \bibinfo {author} {\bibfnamefont {J.~M.}\ \bibnamefont {Conrad}},
  \ and\ \bibinfo {author} {\bibfnamefont {M.~H.}\ \bibnamefont {Shaevitz}},\
  }\href {\doibase 10.1016/j.physrep.2020.08.005} {\bibfield  {journal}
  {\bibinfo  {journal} {Phys. Rept.}\ }\textbf {\bibinfo {volume} {884}},\
  \bibinfo {pages} {1} (\bibinfo {year} {2020})},\ \Eprint
  {http://arxiv.org/abs/1906.00045} {arXiv:1906.00045 [hep-ex]} \BibitemShut
  {NoStop}%
\bibitem [{\citenamefont {B\"oser}\ \emph {et~al.}(2020)\citenamefont
  {B\"oser}, \citenamefont {Buck}, \citenamefont {Giunti}, \citenamefont
  {Lesgourgues}, \citenamefont {Ludhova}, \citenamefont {Mertens},
  \citenamefont {Schukraft},\ and\ \citenamefont {Wurm}}]{Boser:2019rta}%
  \BibitemOpen
  \bibfield  {author} {\bibinfo {author} {\bibfnamefont {S.}~\bibnamefont
  {B\"oser}}, \bibinfo {author} {\bibfnamefont {C.}~\bibnamefont {Buck}},
  \bibinfo {author} {\bibfnamefont {C.}~\bibnamefont {Giunti}}, \bibinfo
  {author} {\bibfnamefont {J.}~\bibnamefont {Lesgourgues}}, \bibinfo {author}
  {\bibfnamefont {L.}~\bibnamefont {Ludhova}}, \bibinfo {author} {\bibfnamefont
  {S.}~\bibnamefont {Mertens}}, \bibinfo {author} {\bibfnamefont
  {A.}~\bibnamefont {Schukraft}}, \ and\ \bibinfo {author} {\bibfnamefont
  {M.}~\bibnamefont {Wurm}},\ }\href {\doibase 10.1016/j.ppnp.2019.103736}
  {\bibfield  {journal} {\bibinfo  {journal} {Prog. Part. Nucl. Phys.}\
  }\textbf {\bibinfo {volume} {111}},\ \bibinfo {pages} {103736} (\bibinfo
  {year} {2020})},\ \Eprint {http://arxiv.org/abs/1906.01739} {arXiv:1906.01739
  [hep-ex]} \BibitemShut {NoStop}%
\bibitem [{\citenamefont {Insler}\ \emph {et~al.}(2010)\citenamefont {Insler}
  \emph {et~al.}}]{Insler:2010jw}%
  \BibitemOpen
  \bibfield  {author} {\bibinfo {author} {\bibfnamefont {J.}~\bibnamefont
  {Insler}} \emph {et~al.} (\bibinfo {collaboration} {CLEO}),\ }\href {\doibase
  10.1103/PhysRevD.81.091101} {\bibfield  {journal} {\bibinfo  {journal} {Phys.
  Rev. D}\ }\textbf {\bibinfo {volume} {81}},\ \bibinfo {pages} {091101}
  (\bibinfo {year} {2010})},\ \Eprint {http://arxiv.org/abs/1003.0417}
  {arXiv:1003.0417 [hep-ex]} \BibitemShut {NoStop}%
\bibitem [{\citenamefont {del Amo~Sanchez}\ \emph {et~al.}(2011)\citenamefont
  {del Amo~Sanchez} \emph {et~al.}}]{delAmoSanchez:2010ac}%
  \BibitemOpen
  \bibfield  {author} {\bibinfo {author} {\bibfnamefont {P.}~\bibnamefont {del
  Amo~Sanchez}} \emph {et~al.} (\bibinfo {collaboration} {BaBar}),\ }\href
  {\doibase 10.1103/PhysRevLett.107.021804} {\bibfield  {journal} {\bibinfo
  {journal} {Phys. Rev. Lett.}\ }\textbf {\bibinfo {volume} {107}},\ \bibinfo
  {pages} {021804} (\bibinfo {year} {2011})},\ \Eprint
  {http://arxiv.org/abs/1007.4646} {arXiv:1007.4646 [hep-ex]} \BibitemShut
  {NoStop}%
\bibitem [{\citenamefont {Seong}\ \emph {et~al.}(2019)\citenamefont {Seong}
  \emph {et~al.}}]{Seong:2018gut}%
  \BibitemOpen
  \bibfield  {author} {\bibinfo {author} {\bibfnamefont {I.~S.}\ \bibnamefont
  {Seong}} \emph {et~al.} (\bibinfo {collaboration} {Belle}),\ }\href {\doibase
  10.1103/PhysRevLett.122.011801} {\bibfield  {journal} {\bibinfo  {journal}
  {Phys. Rev. Lett.}\ }\textbf {\bibinfo {volume} {122}},\ \bibinfo {pages}
  {011801} (\bibinfo {year} {2019})},\ \Eprint
  {http://arxiv.org/abs/1809.05222} {arXiv:1809.05222 [hep-ex]} \BibitemShut
  {NoStop}%
\bibitem [{\citenamefont {Ablikim}\ \emph {et~al.}(2020)\citenamefont {Ablikim}
  \emph {et~al.}}]{Ablikim:2020qtn}%
  \BibitemOpen
  \bibfield  {author} {\bibinfo {author} {\bibfnamefont {M.}~\bibnamefont
  {Ablikim}} \emph {et~al.} (\bibinfo {collaboration} {BESIII}),\ }\href
  {\doibase 10.1103/PhysRevD.101.112005} {\bibfield  {journal} {\bibinfo
  {journal} {Phys. Rev. D}\ }\textbf {\bibinfo {volume} {101}},\ \bibinfo
  {pages} {112005} (\bibinfo {year} {2020})},\ \Eprint
  {http://arxiv.org/abs/2003.05594} {arXiv:2003.05594 [hep-ex]} \BibitemShut
  {NoStop}%
\bibitem [{\citenamefont {Gao}(2014)}]{Gao:2014yga}%
  \BibitemOpen
  \bibfield  {author} {\bibinfo {author} {\bibfnamefont {D.-N.}\ \bibnamefont
  {Gao}},\ }\href {\doibase 10.1103/PhysRevD.90.077501} {\bibfield  {journal}
  {\bibinfo  {journal} {Phys. Rev. D}\ }\textbf {\bibinfo {volume} {90}},\
  \bibinfo {pages} {077501} (\bibinfo {year} {2014})},\ \Eprint
  {http://arxiv.org/abs/1408.4552} {arXiv:1408.4552 [hep-ph]} \BibitemShut
  {NoStop}%
\bibitem [{\citenamefont {Bai}\ \emph {et~al.}(2017)\citenamefont {Bai},
  \citenamefont {Chen},\ and\ \citenamefont {Jia}}]{Bai:2017wle}%
  \BibitemOpen
  \bibfield  {author} {\bibinfo {author} {\bibfnamefont {D.}~\bibnamefont
  {Bai}}, \bibinfo {author} {\bibfnamefont {W.}~\bibnamefont {Chen}}, \ and\
  \bibinfo {author} {\bibfnamefont {Y.}~\bibnamefont {Jia}},\ }\href@noop {} {\
   (\bibinfo {year} {2017})},\ \Eprint {http://arxiv.org/abs/1711.09058}
  {arXiv:1711.09058 [hep-ph]} \BibitemShut {NoStop}%
\bibitem [{\citenamefont {Yeghiyan}(2009)}]{Yeghiyan:2009xc}%
  \BibitemOpen
  \bibfield  {author} {\bibinfo {author} {\bibfnamefont {G.~K.}\ \bibnamefont
  {Yeghiyan}},\ }\href {\doibase 10.1103/PhysRevD.80.115019} {\bibfield
  {journal} {\bibinfo  {journal} {Phys. Rev. D}\ }\textbf {\bibinfo {volume}
  {80}},\ \bibinfo {pages} {115019} (\bibinfo {year} {2009})},\ \Eprint
  {http://arxiv.org/abs/0909.4919} {arXiv:0909.4919 [hep-ph]} \BibitemShut
  {NoStop}%
\bibitem [{\citenamefont {Fernandez}\ \emph {et~al.}(2016)\citenamefont
  {Fernandez}, \citenamefont {Seong},\ and\ \citenamefont
  {Stengel}}]{Fernandez:2015klv}%
  \BibitemOpen
  \bibfield  {author} {\bibinfo {author} {\bibfnamefont {N.}~\bibnamefont
  {Fernandez}}, \bibinfo {author} {\bibfnamefont {I.}~\bibnamefont {Seong}}, \
  and\ \bibinfo {author} {\bibfnamefont {P.}~\bibnamefont {Stengel}},\ }\href
  {\doibase 10.1103/PhysRevD.93.054023} {\bibfield  {journal} {\bibinfo
  {journal} {Phys. Rev. D}\ }\textbf {\bibinfo {volume} {93}},\ \bibinfo
  {pages} {054023} (\bibinfo {year} {2016})},\ \Eprint
  {http://arxiv.org/abs/1511.03728} {arXiv:1511.03728 [hep-ph]} \BibitemShut
  {NoStop}%
\bibitem [{\citenamefont {Liu}\ and\ \citenamefont
  {Zhang}(2019)}]{Liu:2018jdi}%
  \BibitemOpen
  \bibfield  {author} {\bibinfo {author} {\bibfnamefont {Z.}~\bibnamefont
  {Liu}}\ and\ \bibinfo {author} {\bibfnamefont {Y.}~\bibnamefont {Zhang}},\
  }\href {\doibase 10.1103/PhysRevD.99.015004} {\bibfield  {journal} {\bibinfo
  {journal} {Phys. Rev. D}\ }\textbf {\bibinfo {volume} {99}},\ \bibinfo
  {pages} {015004} (\bibinfo {year} {2019})},\ \Eprint
  {http://arxiv.org/abs/1808.00983} {arXiv:1808.00983 [hep-ph]} \BibitemShut
  {NoStop}%
\bibitem [{\citenamefont {Zhang}\ \emph {et~al.}(2019)\citenamefont {Zhang},
  \citenamefont {Zhang}, \citenamefont {Song}, \citenamefont {Pan},
  \citenamefont {Niu},\ and\ \citenamefont {Li}}]{Zhang:2019wnz}%
  \BibitemOpen
  \bibfield  {author} {\bibinfo {author} {\bibfnamefont {Y.}~\bibnamefont
  {Zhang}}, \bibinfo {author} {\bibfnamefont {W.-T.}\ \bibnamefont {Zhang}},
  \bibinfo {author} {\bibfnamefont {M.}~\bibnamefont {Song}}, \bibinfo {author}
  {\bibfnamefont {X.-A.}\ \bibnamefont {Pan}}, \bibinfo {author} {\bibfnamefont
  {Z.-M.}\ \bibnamefont {Niu}}, \ and\ \bibinfo {author} {\bibfnamefont
  {G.}~\bibnamefont {Li}},\ }\href {\doibase 10.1103/PhysRevD.100.115016}
  {\bibfield  {journal} {\bibinfo  {journal} {Phys. Rev. D}\ }\textbf {\bibinfo
  {volume} {100}},\ \bibinfo {pages} {115016} (\bibinfo {year} {2019})},\
  \Eprint {http://arxiv.org/abs/1907.07046} {arXiv:1907.07046 [hep-ph]}
  \BibitemShut {NoStop}%
\bibitem [{\citenamefont {Fayet}(2007)}]{Fayet:2007ua}%
  \BibitemOpen
  \bibfield  {author} {\bibinfo {author} {\bibfnamefont {P.}~\bibnamefont
  {Fayet}},\ }\href {\doibase 10.1103/PhysRevD.75.115017} {\bibfield  {journal}
  {\bibinfo  {journal} {Phys. Rev. D}\ }\textbf {\bibinfo {volume} {75}},\
  \bibinfo {pages} {115017} (\bibinfo {year} {2007})},\ \Eprint
  {http://arxiv.org/abs/hep-ph/0702176} {arXiv:hep-ph/0702176} \BibitemShut
  {NoStop}%
\bibitem [{\citenamefont {McElrath}(2007)}]{McElrath:2007sa}%
  \BibitemOpen
  \bibfield  {author} {\bibinfo {author} {\bibfnamefont {B.}~\bibnamefont
  {McElrath}},\ }\href@noop {} {\bibfield  {journal} {\bibinfo  {journal}
  {eConf}\ }\textbf {\bibinfo {volume} {C070805}},\ \bibinfo {pages} {19}
  (\bibinfo {year} {2007})},\ \Eprint {http://arxiv.org/abs/0712.0016}
  {arXiv:0712.0016 [hep-ph]} \BibitemShut {NoStop}%
\bibitem [{\citenamefont {Fayet}(2010)}]{Fayet:2009tv}%
  \BibitemOpen
  \bibfield  {author} {\bibinfo {author} {\bibfnamefont {P.}~\bibnamefont
  {Fayet}},\ }\href {\doibase 10.1103/PhysRevD.81.054025} {\bibfield  {journal}
  {\bibinfo  {journal} {Phys. Rev. D}\ }\textbf {\bibinfo {volume} {81}},\
  \bibinfo {pages} {054025} (\bibinfo {year} {2010})},\ \Eprint
  {http://arxiv.org/abs/0910.2587} {arXiv:0910.2587 [hep-ph]} \BibitemShut
  {NoStop}%
\bibitem [{\citenamefont {McKeen}(2009)}]{McKeen:2009rm}%
  \BibitemOpen
  \bibfield  {author} {\bibinfo {author} {\bibfnamefont {D.}~\bibnamefont
  {McKeen}},\ }\href {\doibase 10.1103/PhysRevD.79.114001} {\bibfield
  {journal} {\bibinfo  {journal} {Phys. Rev. D}\ }\textbf {\bibinfo {volume}
  {79}},\ \bibinfo {pages} {114001} (\bibinfo {year} {2009})},\ \Eprint
  {http://arxiv.org/abs/0903.4982} {arXiv:0903.4982 [hep-ph]} \BibitemShut
  {NoStop}%
\bibitem [{\citenamefont {Essig}\ \emph {et~al.}(2013)\citenamefont {Essig},
  \citenamefont {Mardon}, \citenamefont {Papucci}, \citenamefont {Volansky},\
  and\ \citenamefont {Zhong}}]{Essig:2013vha}%
  \BibitemOpen
  \bibfield  {author} {\bibinfo {author} {\bibfnamefont {R.}~\bibnamefont
  {Essig}}, \bibinfo {author} {\bibfnamefont {J.}~\bibnamefont {Mardon}},
  \bibinfo {author} {\bibfnamefont {M.}~\bibnamefont {Papucci}}, \bibinfo
  {author} {\bibfnamefont {T.}~\bibnamefont {Volansky}}, \ and\ \bibinfo
  {author} {\bibfnamefont {Y.-M.}\ \bibnamefont {Zhong}},\ }\href {\doibase
  10.1007/JHEP11(2013)167} {\bibfield  {journal} {\bibinfo  {journal} {JHEP}\
  }\textbf {\bibinfo {volume} {11}},\ \bibinfo {pages} {167} (\bibinfo {year}
  {2013})},\ \Eprint {http://arxiv.org/abs/1309.5084} {arXiv:1309.5084
  [hep-ph]} \BibitemShut {NoStop}%
\bibitem [{\citenamefont {Cotta}\ \emph {et~al.}(2014)\citenamefont {Cotta},
  \citenamefont {Rajaraman}, \citenamefont {Tait},\ and\ \citenamefont
  {Wijangco}}]{Cotta:2013jna}%
  \BibitemOpen
  \bibfield  {author} {\bibinfo {author} {\bibfnamefont {R.~C.}\ \bibnamefont
  {Cotta}}, \bibinfo {author} {\bibfnamefont {A.}~\bibnamefont {Rajaraman}},
  \bibinfo {author} {\bibfnamefont {T.~M.~P.}\ \bibnamefont {Tait}}, \ and\
  \bibinfo {author} {\bibfnamefont {A.~M.}\ \bibnamefont {Wijangco}},\ }\href
  {\doibase 10.1103/PhysRevD.90.013020} {\bibfield  {journal} {\bibinfo
  {journal} {Phys. Rev. D}\ }\textbf {\bibinfo {volume} {90}},\ \bibinfo
  {pages} {013020} (\bibinfo {year} {2014})},\ \Eprint
  {http://arxiv.org/abs/1305.6609} {arXiv:1305.6609 [hep-ph]} \BibitemShut
  {NoStop}%
\bibitem [{\citenamefont {Fernandez}\ \emph {et~al.}(2014)\citenamefont
  {Fernandez}, \citenamefont {Kumar}, \citenamefont {Seong},\ and\
  \citenamefont {Stengel}}]{Fernandez:2014eja}%
  \BibitemOpen
  \bibfield  {author} {\bibinfo {author} {\bibfnamefont {N.}~\bibnamefont
  {Fernandez}}, \bibinfo {author} {\bibfnamefont {J.}~\bibnamefont {Kumar}},
  \bibinfo {author} {\bibfnamefont {I.}~\bibnamefont {Seong}}, \ and\ \bibinfo
  {author} {\bibfnamefont {P.}~\bibnamefont {Stengel}},\ }\href {\doibase
  10.1103/PhysRevD.90.015029} {\bibfield  {journal} {\bibinfo  {journal} {Phys.
  Rev. D}\ }\textbf {\bibinfo {volume} {90}},\ \bibinfo {pages} {015029}
  (\bibinfo {year} {2014})},\ \Eprint {http://arxiv.org/abs/1404.6599}
  {arXiv:1404.6599 [hep-ph]} \BibitemShut {NoStop}%
\bibitem [{\citenamefont {Bertuzzo}\ \emph {et~al.}(2017)\citenamefont
  {Bertuzzo}, \citenamefont {Caniu~Barros},\ and\ \citenamefont {Grilli~di
  Cortona}}]{Bertuzzo:2017lwt}%
  \BibitemOpen
  \bibfield  {author} {\bibinfo {author} {\bibfnamefont {E.}~\bibnamefont
  {Bertuzzo}}, \bibinfo {author} {\bibfnamefont {C.~J.}\ \bibnamefont
  {Caniu~Barros}}, \ and\ \bibinfo {author} {\bibfnamefont {G.}~\bibnamefont
  {Grilli~di Cortona}},\ }\href {\doibase 10.1007/JHEP09(2017)116} {\bibfield
  {journal} {\bibinfo  {journal} {JHEP}\ }\textbf {\bibinfo {volume} {09}},\
  \bibinfo {pages} {116} (\bibinfo {year} {2017})},\ \Eprint
  {http://arxiv.org/abs/1707.00725} {arXiv:1707.00725 [hep-ph]} \BibitemShut
  {NoStop}%
\bibitem [{\citenamefont {Bertuzzo}\ and\ \citenamefont
  {Taoso}(2021)}]{Bertuzzo:2020rzo}%
  \BibitemOpen
  \bibfield  {author} {\bibinfo {author} {\bibfnamefont {E.}~\bibnamefont
  {Bertuzzo}}\ and\ \bibinfo {author} {\bibfnamefont {M.}~\bibnamefont
  {Taoso}},\ }\href {\doibase 10.1007/JHEP03(2021)272} {\bibfield  {journal}
  {\bibinfo  {journal} {JHEP}\ }\textbf {\bibinfo {volume} {03}},\ \bibinfo
  {pages} {272} (\bibinfo {year} {2021})},\ \Eprint
  {http://arxiv.org/abs/2011.04735} {arXiv:2011.04735 [hep-ph]} \BibitemShut
  {NoStop}%
\bibitem [{\citenamefont {Chala}\ and\ \citenamefont
  {Titov}(2020)}]{Chala:2020vqp}%
  \BibitemOpen
  \bibfield  {author} {\bibinfo {author} {\bibfnamefont {M.}~\bibnamefont
  {Chala}}\ and\ \bibinfo {author} {\bibfnamefont {A.}~\bibnamefont {Titov}},\
  }\href {\doibase 10.1007/JHEP05(2020)139} {\bibfield  {journal} {\bibinfo
  {journal} {JHEP}\ }\textbf {\bibinfo {volume} {05}},\ \bibinfo {pages} {139}
  (\bibinfo {year} {2020})},\ \Eprint {http://arxiv.org/abs/2001.07732}
  {arXiv:2001.07732 [hep-ph]} \BibitemShut {NoStop}%
\bibitem [{\citenamefont {Li}\ \emph {et~al.}(2020{\natexlab{a}})\citenamefont
  {Li}, \citenamefont {Ma},\ and\ \citenamefont {Schmidt}}]{Li:2020lba}%
  \BibitemOpen
  \bibfield  {author} {\bibinfo {author} {\bibfnamefont {T.}~\bibnamefont
  {Li}}, \bibinfo {author} {\bibfnamefont {X.-D.}\ \bibnamefont {Ma}}, \ and\
  \bibinfo {author} {\bibfnamefont {M.~A.}\ \bibnamefont {Schmidt}},\ }\href
  {\doibase 10.1007/JHEP07(2020)152} {\bibfield  {journal} {\bibinfo  {journal}
  {JHEP}\ }\textbf {\bibinfo {volume} {07}},\ \bibinfo {pages} {152} (\bibinfo
  {year} {2020}{\natexlab{a}})},\ \Eprint {http://arxiv.org/abs/2005.01543}
  {arXiv:2005.01543 [hep-ph]} \BibitemShut {NoStop}%
\bibitem [{\citenamefont {Li}\ \emph {et~al.}(2020{\natexlab{b}})\citenamefont
  {Li}, \citenamefont {Ma},\ and\ \citenamefont {Schmidt}}]{Li:2020wxi}%
  \BibitemOpen
  \bibfield  {author} {\bibinfo {author} {\bibfnamefont {T.}~\bibnamefont
  {Li}}, \bibinfo {author} {\bibfnamefont {X.-D.}\ \bibnamefont {Ma}}, \ and\
  \bibinfo {author} {\bibfnamefont {M.~A.}\ \bibnamefont {Schmidt}},\ }\href
  {\doibase 10.1007/JHEP10(2020)115} {\bibfield  {journal} {\bibinfo  {journal}
  {JHEP}\ }\textbf {\bibinfo {volume} {10}},\ \bibinfo {pages} {115} (\bibinfo
  {year} {2020}{\natexlab{b}})},\ \Eprint {http://arxiv.org/abs/2007.15408}
  {arXiv:2007.15408 [hep-ph]} \BibitemShut {NoStop}%
\bibitem [{\citenamefont {Beltran}\ \emph {et~al.}(2009)\citenamefont
  {Beltran}, \citenamefont {Hooper}, \citenamefont {Kolb},\ and\ \citenamefont
  {Krusberg}}]{Beltran:2008xg}%
  \BibitemOpen
  \bibfield  {author} {\bibinfo {author} {\bibfnamefont {M.}~\bibnamefont
  {Beltran}}, \bibinfo {author} {\bibfnamefont {D.}~\bibnamefont {Hooper}},
  \bibinfo {author} {\bibfnamefont {E.~W.}\ \bibnamefont {Kolb}}, \ and\
  \bibinfo {author} {\bibfnamefont {Z.~C.}\ \bibnamefont {Krusberg}},\ }\href
  {\doibase 10.1103/PhysRevD.80.043509} {\bibfield  {journal} {\bibinfo
  {journal} {Phys. Rev. D}\ }\textbf {\bibinfo {volume} {80}},\ \bibinfo
  {pages} {043509} (\bibinfo {year} {2009})},\ \Eprint
  {http://arxiv.org/abs/0808.3384} {arXiv:0808.3384 [hep-ph]} \BibitemShut
  {NoStop}%
\bibitem [{\citenamefont {Fan}\ \emph {et~al.}(2010)\citenamefont {Fan},
  \citenamefont {Reece},\ and\ \citenamefont {Wang}}]{Fan:2010gt}%
  \BibitemOpen
  \bibfield  {author} {\bibinfo {author} {\bibfnamefont {J.}~\bibnamefont
  {Fan}}, \bibinfo {author} {\bibfnamefont {M.}~\bibnamefont {Reece}}, \ and\
  \bibinfo {author} {\bibfnamefont {L.-T.}\ \bibnamefont {Wang}},\ }\href
  {\doibase 10.1088/1475-7516/2010/11/042} {\bibfield  {journal} {\bibinfo
  {journal} {JCAP}\ }\textbf {\bibinfo {volume} {11}},\ \bibinfo {pages} {042}
  (\bibinfo {year} {2010})},\ \Eprint {http://arxiv.org/abs/1008.1591}
  {arXiv:1008.1591 [hep-ph]} \BibitemShut {NoStop}%
\bibitem [{\citenamefont {Goodman}\ \emph {et~al.}(2011)\citenamefont
  {Goodman}, \citenamefont {Ibe}, \citenamefont {Rajaraman}, \citenamefont
  {Shepherd}, \citenamefont {Tait},\ and\ \citenamefont {Yu}}]{Goodman:2010qn}%
  \BibitemOpen
  \bibfield  {author} {\bibinfo {author} {\bibfnamefont {J.}~\bibnamefont
  {Goodman}}, \bibinfo {author} {\bibfnamefont {M.}~\bibnamefont {Ibe}},
  \bibinfo {author} {\bibfnamefont {A.}~\bibnamefont {Rajaraman}}, \bibinfo
  {author} {\bibfnamefont {W.}~\bibnamefont {Shepherd}}, \bibinfo {author}
  {\bibfnamefont {T.~M.~P.}\ \bibnamefont {Tait}}, \ and\ \bibinfo {author}
  {\bibfnamefont {H.-B.}\ \bibnamefont {Yu}},\ }\href {\doibase
  10.1016/j.nuclphysb.2010.10.022} {\bibfield  {journal} {\bibinfo  {journal}
  {Nucl. Phys. B}\ }\textbf {\bibinfo {volume} {844}},\ \bibinfo {pages} {55}
  (\bibinfo {year} {2011})},\ \Eprint {http://arxiv.org/abs/1009.0008}
  {arXiv:1009.0008 [hep-ph]} \BibitemShut {NoStop}%
\bibitem [{\citenamefont {Bal\'azs}\ \emph {et~al.}(2014)\citenamefont
  {Bal\'azs}, \citenamefont {Li},\ and\ \citenamefont
  {Newstead}}]{Balazs:2014rsa}%
  \BibitemOpen
  \bibfield  {author} {\bibinfo {author} {\bibfnamefont {C.}~\bibnamefont
  {Bal\'azs}}, \bibinfo {author} {\bibfnamefont {T.}~\bibnamefont {Li}}, \ and\
  \bibinfo {author} {\bibfnamefont {J.~L.}\ \bibnamefont {Newstead}},\ }\href
  {\doibase 10.1007/JHEP08(2014)061} {\bibfield  {journal} {\bibinfo  {journal}
  {JHEP}\ }\textbf {\bibinfo {volume} {08}},\ \bibinfo {pages} {061} (\bibinfo
  {year} {2014})},\ \Eprint {http://arxiv.org/abs/1403.5829} {arXiv:1403.5829
  [hep-ph]} \BibitemShut {NoStop}%
\bibitem [{\citenamefont {De~Simone}\ and\ \citenamefont
  {Jacques}(2016)}]{DeSimone:2016fbz}%
  \BibitemOpen
  \bibfield  {author} {\bibinfo {author} {\bibfnamefont {A.}~\bibnamefont
  {De~Simone}}\ and\ \bibinfo {author} {\bibfnamefont {T.}~\bibnamefont
  {Jacques}},\ }\href {\doibase 10.1140/epjc/s10052-016-4208-4} {\bibfield
  {journal} {\bibinfo  {journal} {Eur. Phys. J. C}\ }\textbf {\bibinfo {volume}
  {76}},\ \bibinfo {pages} {367} (\bibinfo {year} {2016})},\ \Eprint
  {http://arxiv.org/abs/1603.08002} {arXiv:1603.08002 [hep-ph]} \BibitemShut
  {NoStop}%
\bibitem [{\citenamefont {Brod}\ \emph {et~al.}(2018)\citenamefont {Brod},
  \citenamefont {Gootjes-Dreesbach}, \citenamefont {Tammaro},\ and\
  \citenamefont {Zupan}}]{Brod:2017bsw}%
  \BibitemOpen
  \bibfield  {author} {\bibinfo {author} {\bibfnamefont {J.}~\bibnamefont
  {Brod}}, \bibinfo {author} {\bibfnamefont {A.}~\bibnamefont
  {Gootjes-Dreesbach}}, \bibinfo {author} {\bibfnamefont {M.}~\bibnamefont
  {Tammaro}}, \ and\ \bibinfo {author} {\bibfnamefont {J.}~\bibnamefont
  {Zupan}},\ }\href {\doibase 10.1007/JHEP10(2018)065} {\bibfield  {journal}
  {\bibinfo  {journal} {JHEP}\ }\textbf {\bibinfo {volume} {10}},\ \bibinfo
  {pages} {065} (\bibinfo {year} {2018})},\ \Eprint
  {http://arxiv.org/abs/1710.10218} {arXiv:1710.10218 [hep-ph]} \BibitemShut
  {NoStop}%
\bibitem [{\citenamefont {Fitzpatrick}\ \emph {et~al.}(2013)\citenamefont
  {Fitzpatrick}, \citenamefont {Haxton}, \citenamefont {Katz}, \citenamefont
  {Lubbers},\ and\ \citenamefont {Xu}}]{Fitzpatrick:2012ix}%
  \BibitemOpen
  \bibfield  {author} {\bibinfo {author} {\bibfnamefont {A.~L.}\ \bibnamefont
  {Fitzpatrick}}, \bibinfo {author} {\bibfnamefont {W.}~\bibnamefont {Haxton}},
  \bibinfo {author} {\bibfnamefont {E.}~\bibnamefont {Katz}}, \bibinfo {author}
  {\bibfnamefont {N.}~\bibnamefont {Lubbers}}, \ and\ \bibinfo {author}
  {\bibfnamefont {Y.}~\bibnamefont {Xu}},\ }\href {\doibase
  10.1088/1475-7516/2013/02/004} {\bibfield  {journal} {\bibinfo  {journal}
  {JCAP}\ }\textbf {\bibinfo {volume} {02}},\ \bibinfo {pages} {004} (\bibinfo
  {year} {2013})},\ \Eprint {http://arxiv.org/abs/1203.3542} {arXiv:1203.3542
  [hep-ph]} \BibitemShut {NoStop}%
\bibitem [{\citenamefont {Fitzpatrick}\ \emph {et~al.}(2012)\citenamefont
  {Fitzpatrick}, \citenamefont {Haxton}, \citenamefont {Katz}, \citenamefont
  {Lubbers},\ and\ \citenamefont {Xu}}]{Fitzpatrick:2012ib}%
  \BibitemOpen
  \bibfield  {author} {\bibinfo {author} {\bibfnamefont {A.~L.}\ \bibnamefont
  {Fitzpatrick}}, \bibinfo {author} {\bibfnamefont {W.}~\bibnamefont {Haxton}},
  \bibinfo {author} {\bibfnamefont {E.}~\bibnamefont {Katz}}, \bibinfo {author}
  {\bibfnamefont {N.}~\bibnamefont {Lubbers}}, \ and\ \bibinfo {author}
  {\bibfnamefont {Y.}~\bibnamefont {Xu}},\ }\href@noop {} {\  (\bibinfo {year}
  {2012})},\ \Eprint {http://arxiv.org/abs/1211.2818} {arXiv:1211.2818
  [hep-ph]} \BibitemShut {NoStop}%
\bibitem [{\citenamefont {Cirelli}\ \emph {et~al.}(2013)\citenamefont
  {Cirelli}, \citenamefont {Del~Nobile},\ and\ \citenamefont
  {Panci}}]{DelNobile:2013sia}%
  \BibitemOpen
  \bibfield  {author} {\bibinfo {author} {\bibfnamefont {M.}~\bibnamefont
  {Cirelli}}, \bibinfo {author} {\bibfnamefont {E.}~\bibnamefont {Del~Nobile}},
  \ and\ \bibinfo {author} {\bibfnamefont {P.}~\bibnamefont {Panci}},\ }\href
  {\doibase 10.1088/1475-7516/2013/10/019} {\bibfield  {journal} {\bibinfo
  {journal} {JCAP}\ }\textbf {\bibinfo {volume} {10}},\ \bibinfo {pages} {019}
  (\bibinfo {year} {2013})},\ \Eprint {http://arxiv.org/abs/1307.5955}
  {arXiv:1307.5955 [hep-ph]} \BibitemShut {NoStop}%
\bibitem [{\citenamefont {Bishara}\ \emph {et~al.}(2017)\citenamefont
  {Bishara}, \citenamefont {Brod}, \citenamefont {Grinstein},\ and\
  \citenamefont {Zupan}}]{Bishara:2017pfq}%
  \BibitemOpen
  \bibfield  {author} {\bibinfo {author} {\bibfnamefont {F.}~\bibnamefont
  {Bishara}}, \bibinfo {author} {\bibfnamefont {J.}~\bibnamefont {Brod}},
  \bibinfo {author} {\bibfnamefont {B.}~\bibnamefont {Grinstein}}, \ and\
  \bibinfo {author} {\bibfnamefont {J.}~\bibnamefont {Zupan}},\ }\href
  {\doibase 10.1007/JHEP11(2017)059} {\bibfield  {journal} {\bibinfo  {journal}
  {JHEP}\ }\textbf {\bibinfo {volume} {11}},\ \bibinfo {pages} {059} (\bibinfo
  {year} {2017})},\ \Eprint {http://arxiv.org/abs/1707.06998} {arXiv:1707.06998
  [hep-ph]} \BibitemShut {NoStop}%
\bibitem [{\citenamefont {Bischer}\ \emph {et~al.}(2021)\citenamefont
  {Bischer}, \citenamefont {Plehn},\ and\ \citenamefont
  {Rodejohann}}]{Bischer:2020sop}%
  \BibitemOpen
  \bibfield  {author} {\bibinfo {author} {\bibfnamefont {I.}~\bibnamefont
  {Bischer}}, \bibinfo {author} {\bibfnamefont {T.}~\bibnamefont {Plehn}}, \
  and\ \bibinfo {author} {\bibfnamefont {W.}~\bibnamefont {Rodejohann}},\
  }\href {\doibase 10.21468/SciPostPhys.10.2.039} {\bibfield  {journal}
  {\bibinfo  {journal} {SciPost Phys.}\ }\textbf {\bibinfo {volume} {10}},\
  \bibinfo {pages} {039} (\bibinfo {year} {2021})},\ \Eprint
  {http://arxiv.org/abs/2008.04718} {arXiv:2008.04718 [hep-ph]} \BibitemShut
  {NoStop}%
\bibitem [{\citenamefont {Peng}()}]{STCF0}%
  \BibitemOpen
  \bibfield  {author} {\bibinfo {author} {\bibfnamefont {H.-p.}\ \bibnamefont
  {Peng}},\ }in\ \href@noop {} {\emph {\bibinfo {booktitle} {{talk at
  Charm2018, Novosibirsk, Russia, May 21 - 25, 2018}}}}\BibitemShut {NoStop}%
\bibitem [{\citenamefont {Altmannshofer}\ \emph {et~al.}(2019)\citenamefont
  {Altmannshofer} \emph {et~al.}}]{Belle-II:2018jsg}%
  \BibitemOpen
  \bibfield  {author} {\bibinfo {author} {\bibfnamefont {W.}~\bibnamefont
  {Altmannshofer}} \emph {et~al.} (\bibinfo {collaboration} {Belle-II}),\
  }\href {\doibase 10.1093/ptep/ptz106} {\bibfield  {journal} {\bibinfo
  {journal} {PTEP}\ }\textbf {\bibinfo {volume} {2019}},\ \bibinfo {pages}
  {123C01} (\bibinfo {year} {2019})},\ \bibinfo {note} {[Erratum: PTEP 2020,
  029201 (2020)]},\ \Eprint {http://arxiv.org/abs/1808.10567} {arXiv:1808.10567
  [hep-ex]} \BibitemShut {NoStop}%
\bibitem [{\citenamefont {Aaij}\ \emph {et~al.}(2018)\citenamefont {Aaij} \emph
  {et~al.}}]{LHCb:2018roe}%
  \BibitemOpen
  \bibfield  {author} {\bibinfo {author} {\bibfnamefont {R.}~\bibnamefont
  {Aaij}} \emph {et~al.} (\bibinfo {collaboration} {LHCb}),\ }\href@noop {} {\
  (\bibinfo {year} {2018})},\ \Eprint {http://arxiv.org/abs/1808.08865}
  {arXiv:1808.08865 [hep-ex]} \BibitemShut {NoStop}%
\bibitem [{\citenamefont {Fayet}(1979)}]{Fayet:1979qi}%
  \BibitemOpen
  \bibfield  {author} {\bibinfo {author} {\bibfnamefont {P.}~\bibnamefont
  {Fayet}},\ }\href {\doibase 10.1016/0370-2693(79)91230-9} {\bibfield
  {journal} {\bibinfo  {journal} {Phys. Lett. B}\ }\textbf {\bibinfo {volume}
  {84}},\ \bibinfo {pages} {421} (\bibinfo {year} {1979})}\BibitemShut
  {NoStop}%
\bibitem [{\citenamefont {Fayet}\ and\ \citenamefont
  {Kaplan}(1991)}]{Fayet:1991ux}%
  \BibitemOpen
  \bibfield  {author} {\bibinfo {author} {\bibfnamefont {P.}~\bibnamefont
  {Fayet}}\ and\ \bibinfo {author} {\bibfnamefont {J.}~\bibnamefont {Kaplan}},\
  }\href {\doibase 10.1016/0370-2693(91)91477-D} {\bibfield  {journal}
  {\bibinfo  {journal} {Phys. Lett. B}\ }\textbf {\bibinfo {volume} {269}},\
  \bibinfo {pages} {213} (\bibinfo {year} {1991})}\BibitemShut {NoStop}%
\bibitem [{\citenamefont {McElrath}(2005)}]{McElrath:2005bp}%
  \BibitemOpen
  \bibfield  {author} {\bibinfo {author} {\bibfnamefont {B.}~\bibnamefont
  {McElrath}},\ }\href {\doibase 10.1103/PhysRevD.72.103508} {\bibfield
  {journal} {\bibinfo  {journal} {Phys. Rev. D}\ }\textbf {\bibinfo {volume}
  {72}},\ \bibinfo {pages} {103508} (\bibinfo {year} {2005})},\ \Eprint
  {http://arxiv.org/abs/hep-ph/0506151} {arXiv:hep-ph/0506151} \BibitemShut
  {NoStop}%
\bibitem [{\citenamefont {Chang}\ \emph {et~al.}(1998)\citenamefont {Chang},
  \citenamefont {Lebedev},\ and\ \citenamefont {Ng}}]{Chang:1997tq}%
  \BibitemOpen
  \bibfield  {author} {\bibinfo {author} {\bibfnamefont {L.~N.}\ \bibnamefont
  {Chang}}, \bibinfo {author} {\bibfnamefont {O.}~\bibnamefont {Lebedev}}, \
  and\ \bibinfo {author} {\bibfnamefont {J.~N.}\ \bibnamefont {Ng}},\ }\href
  {\doibase 10.1016/S0370-2693(98)01147-2} {\bibfield  {journal} {\bibinfo
  {journal} {Phys. Lett. B}\ }\textbf {\bibinfo {volume} {441}},\ \bibinfo
  {pages} {419} (\bibinfo {year} {1998})},\ \Eprint
  {http://arxiv.org/abs/hep-ph/9806487} {arXiv:hep-ph/9806487} \BibitemShut
  {NoStop}%
\bibitem [{\citenamefont {Ablikim}\ \emph {et~al.}(2008)\citenamefont {Ablikim}
  \emph {et~al.}}]{Ablikim:2007ek}%
  \BibitemOpen
  \bibfield  {author} {\bibinfo {author} {\bibfnamefont {M.}~\bibnamefont
  {Ablikim}} \emph {et~al.} (\bibinfo {collaboration} {BES}),\ }\href {\doibase
  10.1103/PhysRevLett.100.192001} {\bibfield  {journal} {\bibinfo  {journal}
  {Phys. Rev. Lett.}\ }\textbf {\bibinfo {volume} {100}},\ \bibinfo {pages}
  {192001} (\bibinfo {year} {2008})},\ \Eprint {http://arxiv.org/abs/0710.0039}
  {arXiv:0710.0039 [hep-ex]} \BibitemShut {NoStop}%
\bibitem [{\citenamefont {Zyla}\ \emph {et~al.}(2020)\citenamefont {Zyla} \emph
  {et~al.}}]{Zyla:2020zbs}%
  \BibitemOpen
  \bibfield  {author} {\bibinfo {author} {\bibfnamefont {P.~A.}\ \bibnamefont
  {Zyla}} \emph {et~al.} (\bibinfo {collaboration} {Particle Data Group}),\
  }\href {\doibase 10.1093/ptep/ptaa104} {\bibfield  {journal} {\bibinfo
  {journal} {PTEP}\ }\textbf {\bibinfo {volume} {2020}},\ \bibinfo {pages}
  {083C01} (\bibinfo {year} {2020})}\BibitemShut {NoStop}%
\bibitem [{\citenamefont {Liao}\ \emph {et~al.}(2020)\citenamefont {Liao},
  \citenamefont {Ma},\ and\ \citenamefont {Wang}}]{Liao:2020zyx}%
  \BibitemOpen
  \bibfield  {author} {\bibinfo {author} {\bibfnamefont {Y.}~\bibnamefont
  {Liao}}, \bibinfo {author} {\bibfnamefont {X.-D.}\ \bibnamefont {Ma}}, \ and\
  \bibinfo {author} {\bibfnamefont {Q.-Y.}\ \bibnamefont {Wang}},\ }\href
  {\doibase 10.1007/JHEP08(2020)162} {\bibfield  {journal} {\bibinfo  {journal}
  {JHEP}\ }\textbf {\bibinfo {volume} {08}},\ \bibinfo {pages} {162} (\bibinfo
  {year} {2020})},\ \Eprint {http://arxiv.org/abs/2005.08013} {arXiv:2005.08013
  [hep-ph]} \BibitemShut {NoStop}%
\bibitem [{\citenamefont {Jenkins}\ \emph {et~al.}(2018)\citenamefont
  {Jenkins}, \citenamefont {Manohar},\ and\ \citenamefont
  {Stoffer}}]{Jenkins:2017jig}%
  \BibitemOpen
  \bibfield  {author} {\bibinfo {author} {\bibfnamefont {E.~E.}\ \bibnamefont
  {Jenkins}}, \bibinfo {author} {\bibfnamefont {A.~V.}\ \bibnamefont
  {Manohar}}, \ and\ \bibinfo {author} {\bibfnamefont {P.}~\bibnamefont
  {Stoffer}},\ }\href {\doibase 10.1007/JHEP03(2018)016} {\bibfield  {journal}
  {\bibinfo  {journal} {JHEP}\ }\textbf {\bibinfo {volume} {03}},\ \bibinfo
  {pages} {016} (\bibinfo {year} {2018})},\ \Eprint
  {http://arxiv.org/abs/1709.04486} {arXiv:1709.04486 [hep-ph]} \BibitemShut
  {NoStop}%
\bibitem [{\citenamefont {Appelquist}\ and\ \citenamefont
  {Politzer}(1975)}]{Appelquist:1974zd}%
  \BibitemOpen
  \bibfield  {author} {\bibinfo {author} {\bibfnamefont {T.}~\bibnamefont
  {Appelquist}}\ and\ \bibinfo {author} {\bibfnamefont {H.~D.}\ \bibnamefont
  {Politzer}},\ }\href {\doibase 10.1103/PhysRevLett.34.43} {\bibfield
  {journal} {\bibinfo  {journal} {Phys. Rev. Lett.}\ }\textbf {\bibinfo
  {volume} {34}},\ \bibinfo {pages} {43} (\bibinfo {year} {1975})}\BibitemShut
  {NoStop}%
\bibitem [{\citenamefont {De~Rujula}\ and\ \citenamefont
  {Glashow}(1975)}]{DeRujula:1974rkb}%
  \BibitemOpen
  \bibfield  {author} {\bibinfo {author} {\bibfnamefont {A.}~\bibnamefont
  {De~Rujula}}\ and\ \bibinfo {author} {\bibfnamefont {S.~L.}\ \bibnamefont
  {Glashow}},\ }\href {\doibase 10.1103/PhysRevLett.34.46} {\bibfield
  {journal} {\bibinfo  {journal} {Phys. Rev. Lett.}\ }\textbf {\bibinfo
  {volume} {34}},\ \bibinfo {pages} {46} (\bibinfo {year} {1975})}\BibitemShut
  {NoStop}%
\bibitem [{\citenamefont {Kuhn}\ \emph {et~al.}(1979)\citenamefont {Kuhn},
  \citenamefont {Kaplan},\ and\ \citenamefont {Safiani}}]{Kuhn:1979bb}%
  \BibitemOpen
  \bibfield  {author} {\bibinfo {author} {\bibfnamefont {J.~H.}\ \bibnamefont
  {Kuhn}}, \bibinfo {author} {\bibfnamefont {J.}~\bibnamefont {Kaplan}}, \ and\
  \bibinfo {author} {\bibfnamefont {E.~G.~O.}\ \bibnamefont {Safiani}},\ }\href
  {\doibase 10.1016/0550-3213(79)90055-5} {\bibfield  {journal} {\bibinfo
  {journal} {Nucl. Phys. B}\ }\textbf {\bibinfo {volume} {157}},\ \bibinfo
  {pages} {125} (\bibinfo {year} {1979})}\BibitemShut {NoStop}%
\bibitem [{\citenamefont {Keung}(1981)}]{Keung:1980ev}%
  \BibitemOpen
  \bibfield  {author} {\bibinfo {author} {\bibfnamefont {W.-Y.}\ \bibnamefont
  {Keung}},\ }\href {\doibase 10.1103/PhysRevD.23.2072} {\bibfield  {journal}
  {\bibinfo  {journal} {Phys. Rev. D}\ }\textbf {\bibinfo {volume} {23}},\
  \bibinfo {pages} {2072} (\bibinfo {year} {1981})}\BibitemShut {NoStop}%
\bibitem [{\citenamefont {Berger}\ and\ \citenamefont
  {Jones}(1981)}]{Berger:1980ni}%
  \BibitemOpen
  \bibfield  {author} {\bibinfo {author} {\bibfnamefont {E.~L.}\ \bibnamefont
  {Berger}}\ and\ \bibinfo {author} {\bibfnamefont {D.~L.}\ \bibnamefont
  {Jones}},\ }\href {\doibase 10.1103/PhysRevD.23.1521} {\bibfield  {journal}
  {\bibinfo  {journal} {Phys. Rev. D}\ }\textbf {\bibinfo {volume} {23}},\
  \bibinfo {pages} {1521} (\bibinfo {year} {1981})}\BibitemShut {NoStop}%
\bibitem [{\citenamefont {Clavelli}(1982)}]{Clavelli:1982hp}%
  \BibitemOpen
  \bibfield  {author} {\bibinfo {author} {\bibfnamefont {L.}~\bibnamefont
  {Clavelli}},\ }\href {\doibase 10.1103/PhysRevD.26.1610} {\bibfield
  {journal} {\bibinfo  {journal} {Phys. Rev. D}\ }\textbf {\bibinfo {volume}
  {26}},\ \bibinfo {pages} {1610} (\bibinfo {year} {1982})}\BibitemShut
  {NoStop}%
\bibitem [{\citenamefont {Clavelli}\ \emph {et~al.}(2001)\citenamefont
  {Clavelli}, \citenamefont {Gajdosik},\ and\ \citenamefont
  {Perevalova}}]{Clavelli:2001zi}%
  \BibitemOpen
  \bibfield  {author} {\bibinfo {author} {\bibfnamefont {L.}~\bibnamefont
  {Clavelli}}, \bibinfo {author} {\bibfnamefont {T.}~\bibnamefont {Gajdosik}},
  \ and\ \bibinfo {author} {\bibfnamefont {I.}~\bibnamefont {Perevalova}},\
  }\href {\doibase 10.1016/S0370-2693(01)01359-4} {\bibfield  {journal}
  {\bibinfo  {journal} {Phys. Lett. B}\ }\textbf {\bibinfo {volume} {523}},\
  \bibinfo {pages} {249} (\bibinfo {year} {2001})},\ \Eprint
  {http://arxiv.org/abs/hep-ph/0110076} {arXiv:hep-ph/0110076} \BibitemShut
  {NoStop}%
\bibitem [{\citenamefont {Clavelli}\ \emph {et~al.}(2002)\citenamefont
  {Clavelli}, \citenamefont {Coulter},\ and\ \citenamefont
  {Gajdosik}}]{Clavelli:2001gb}%
  \BibitemOpen
  \bibfield  {author} {\bibinfo {author} {\bibfnamefont {L.}~\bibnamefont
  {Clavelli}}, \bibinfo {author} {\bibfnamefont {P.}~\bibnamefont {Coulter}}, \
  and\ \bibinfo {author} {\bibfnamefont {T.}~\bibnamefont {Gajdosik}},\ }\href
  {\doibase 10.1016/S0370-2693(01)01516-7} {\bibfield  {journal} {\bibinfo
  {journal} {Phys. Lett. B}\ }\textbf {\bibinfo {volume} {526}},\ \bibinfo
  {pages} {360} (\bibinfo {year} {2002})},\ \Eprint
  {http://arxiv.org/abs/hep-ph/0111250} {arXiv:hep-ph/0111250} \BibitemShut
  {NoStop}%
\bibitem [{\citenamefont {Hao}\ \emph {et~al.}(2007)\citenamefont {Hao},
  \citenamefont {Jia}, \citenamefont {Qiao},\ and\ \citenamefont
  {Sun}}]{Hao:2006nf}%
  \BibitemOpen
  \bibfield  {author} {\bibinfo {author} {\bibfnamefont {G.}~\bibnamefont
  {Hao}}, \bibinfo {author} {\bibfnamefont {Y.}~\bibnamefont {Jia}}, \bibinfo
  {author} {\bibfnamefont {C.-F.}\ \bibnamefont {Qiao}}, \ and\ \bibinfo
  {author} {\bibfnamefont {P.}~\bibnamefont {Sun}},\ }\href {\doibase
  10.1088/1126-6708/2007/02/057} {\bibfield  {journal} {\bibinfo  {journal}
  {JHEP}\ }\textbf {\bibinfo {volume} {02}},\ \bibinfo {pages} {057} (\bibinfo
  {year} {2007})},\ \Eprint {http://arxiv.org/abs/hep-ph/0612173}
  {arXiv:hep-ph/0612173} \BibitemShut {NoStop}%
\bibitem [{\citenamefont {Ball}\ \emph {et~al.}(2007)\citenamefont {Ball},
  \citenamefont {Jones},\ and\ \citenamefont {Zwicky}}]{Ball:2006eu}%
  \BibitemOpen
  \bibfield  {author} {\bibinfo {author} {\bibfnamefont {P.}~\bibnamefont
  {Ball}}, \bibinfo {author} {\bibfnamefont {G.~W.}\ \bibnamefont {Jones}}, \
  and\ \bibinfo {author} {\bibfnamefont {R.}~\bibnamefont {Zwicky}},\ }\href
  {\doibase 10.1103/PhysRevD.75.054004} {\bibfield  {journal} {\bibinfo
  {journal} {Phys. Rev. D}\ }\textbf {\bibinfo {volume} {75}},\ \bibinfo
  {pages} {054004} (\bibinfo {year} {2007})},\ \Eprint
  {http://arxiv.org/abs/hep-ph/0612081} {arXiv:hep-ph/0612081} \BibitemShut
  {NoStop}%
\bibitem [{\citenamefont {Cheng}\ \emph {et~al.}(2013)\citenamefont {Cheng},
  \citenamefont {Chua}, \citenamefont {Yang},\ and\ \citenamefont
  {Zhang}}]{Cheng:2013fba}%
  \BibitemOpen
  \bibfield  {author} {\bibinfo {author} {\bibfnamefont {H.-Y.}\ \bibnamefont
  {Cheng}}, \bibinfo {author} {\bibfnamefont {C.-K.}\ \bibnamefont {Chua}},
  \bibinfo {author} {\bibfnamefont {K.-C.}\ \bibnamefont {Yang}}, \ and\
  \bibinfo {author} {\bibfnamefont {Z.-Q.}\ \bibnamefont {Zhang}},\ }\href
  {\doibase 10.1103/PhysRevD.87.114001} {\bibfield  {journal} {\bibinfo
  {journal} {Phys.\ Rev.\ D}\ }\textbf {\bibinfo {volume} {87}},\ \bibinfo
  {pages} {114001} (\bibinfo {year} {2013})},\ \Eprint
  {http://arxiv.org/abs/1303.4403} {arXiv:1303.4403 [hep-ph]} \BibitemShut
  {NoStop}%
\bibitem [{\citenamefont {Barger}\ and\ \citenamefont {Phillips}(1996)}]{CP}%
  \BibitemOpen
  \bibfield  {author} {\bibinfo {author} {\bibfnamefont {V.~D.}\ \bibnamefont
  {Barger}}\ and\ \bibinfo {author} {\bibfnamefont {R.~J.}\ \bibnamefont
  {Phillips}},\ }\href@noop {} {\bibfield  {journal} {\bibinfo  {journal}
  {Collider Physics (Updated Edition), Westview Press}\ } (\bibinfo {year}
  {1996})}\BibitemShut {NoStop}%
\bibitem [{\citenamefont {Read}(2000)}]{Read:2000ru}%
  \BibitemOpen
  \bibfield  {author} {\bibinfo {author} {\bibfnamefont {A.~L.}\ \bibnamefont
  {Read}},\ }in\ \href@noop {} {\emph {\bibinfo {booktitle} {{Workshop on
  Confidence Limits}}}}\ (\bibinfo {year} {2000})\ pp.\ \bibinfo {pages}
  {81--101}\BibitemShut {NoStop}%
\bibitem [{\citenamefont {Read}(2002)}]{Read:2002hq}%
  \BibitemOpen
  \bibfield  {author} {\bibinfo {author} {\bibfnamefont {A.~L.}\ \bibnamefont
  {Read}},\ }\href {\doibase 10.1088/0954-3899/28/10/313} {\bibfield  {journal}
  {\bibinfo  {journal} {J. Phys. G}\ }\textbf {\bibinfo {volume} {28}},\
  \bibinfo {pages} {2693} (\bibinfo {year} {2002})}\BibitemShut {NoStop}%
\bibitem [{\citenamefont {Cowan}\ \emph {et~al.}(2011)\citenamefont {Cowan},
  \citenamefont {Cranmer}, \citenamefont {Gross},\ and\ \citenamefont
  {Vitells}}]{Cowan:2010js}%
  \BibitemOpen
  \bibfield  {author} {\bibinfo {author} {\bibfnamefont {G.}~\bibnamefont
  {Cowan}}, \bibinfo {author} {\bibfnamefont {K.}~\bibnamefont {Cranmer}},
  \bibinfo {author} {\bibfnamefont {E.}~\bibnamefont {Gross}}, \ and\ \bibinfo
  {author} {\bibfnamefont {O.}~\bibnamefont {Vitells}},\ }\href {\doibase
  10.1140/epjc/s10052-011-1554-0} {\bibfield  {journal} {\bibinfo  {journal}
  {Eur. Phys. J. C}\ }\textbf {\bibinfo {volume} {71}},\ \bibinfo {pages}
  {1554} (\bibinfo {year} {2011})},\ \bibinfo {note} {[Erratum: Eur.Phys.J.C
  73, 2501 (2013)]},\ \Eprint {http://arxiv.org/abs/1007.1727} {arXiv:1007.1727
  [physics.data-an]} \BibitemShut {NoStop}%
\bibitem [{\citenamefont {Liu}\ \emph {et~al.}(2019)\citenamefont {Liu} \emph
  {et~al.}}]{Liu:2019kzq}%
  \BibitemOpen
  \bibfield  {author} {\bibinfo {author} {\bibfnamefont {Z.~Z.}\ \bibnamefont
  {Liu}} \emph {et~al.} (\bibinfo {collaboration} {CDEX}),\ }\href {\doibase
  10.1103/PhysRevLett.123.161301} {\bibfield  {journal} {\bibinfo  {journal}
  {Phys. Rev. Lett.}\ }\textbf {\bibinfo {volume} {123}},\ \bibinfo {pages}
  {161301} (\bibinfo {year} {2019})},\ \Eprint
  {http://arxiv.org/abs/1905.00354} {arXiv:1905.00354 [hep-ex]} \BibitemShut
  {NoStop}%
\bibitem [{\citenamefont {Agnese}\ \emph {et~al.}(2018)\citenamefont {Agnese}
  \emph {et~al.}}]{Agnese:2017jvy}%
  \BibitemOpen
  \bibfield  {author} {\bibinfo {author} {\bibfnamefont {R.}~\bibnamefont
  {Agnese}} \emph {et~al.} (\bibinfo {collaboration} {SuperCDMS}),\ }\href
  {\doibase 10.1103/PhysRevD.97.022002} {\bibfield  {journal} {\bibinfo
  {journal} {Phys. Rev. D}\ }\textbf {\bibinfo {volume} {97}},\ \bibinfo
  {pages} {022002} (\bibinfo {year} {2018})},\ \Eprint
  {http://arxiv.org/abs/1707.01632} {arXiv:1707.01632 [astro-ph.CO]}
  \BibitemShut {NoStop}%
\bibitem [{\citenamefont {Abdelhameed}\ \emph {et~al.}(2019)\citenamefont
  {Abdelhameed} \emph {et~al.}}]{Abdelhameed:2019hmk}%
  \BibitemOpen
  \bibfield  {author} {\bibinfo {author} {\bibfnamefont {A.~H.}\ \bibnamefont
  {Abdelhameed}} \emph {et~al.} (\bibinfo {collaboration} {CRESST}),\ }\href
  {\doibase 10.1103/PhysRevD.100.102002} {\bibfield  {journal} {\bibinfo
  {journal} {Phys. Rev. D}\ }\textbf {\bibinfo {volume} {100}},\ \bibinfo
  {pages} {102002} (\bibinfo {year} {2019})},\ \Eprint
  {http://arxiv.org/abs/1904.00498} {arXiv:1904.00498 [astro-ph.CO]}
  \BibitemShut {NoStop}%
\bibitem [{\citenamefont {Agnes}\ \emph {et~al.}(2018)\citenamefont {Agnes}
  \emph {et~al.}}]{Agnes:2018ves}%
  \BibitemOpen
  \bibfield  {author} {\bibinfo {author} {\bibfnamefont {P.}~\bibnamefont
  {Agnes}} \emph {et~al.} (\bibinfo {collaboration} {DarkSide}),\ }\href
  {\doibase 10.1103/PhysRevLett.121.081307} {\bibfield  {journal} {\bibinfo
  {journal} {Phys. Rev. Lett.}\ }\textbf {\bibinfo {volume} {121}},\ \bibinfo
  {pages} {081307} (\bibinfo {year} {2018})},\ \Eprint
  {http://arxiv.org/abs/1802.06994} {arXiv:1802.06994 [astro-ph.HE]}
  \BibitemShut {NoStop}%
\bibitem [{\citenamefont {Jiang}\ \emph {et~al.}(2018)\citenamefont {Jiang}
  \emph {et~al.}}]{Jiang:2018pic}%
  \BibitemOpen
  \bibfield  {author} {\bibinfo {author} {\bibfnamefont {H.}~\bibnamefont
  {Jiang}} \emph {et~al.} (\bibinfo {collaboration} {CDEX}),\ }\href {\doibase
  10.1103/PhysRevLett.120.241301} {\bibfield  {journal} {\bibinfo  {journal}
  {Phys. Rev. Lett.}\ }\textbf {\bibinfo {volume} {120}},\ \bibinfo {pages}
  {241301} (\bibinfo {year} {2018})},\ \Eprint
  {http://arxiv.org/abs/1802.09016} {arXiv:1802.09016 [hep-ex]} \BibitemShut
  {NoStop}%
\bibitem [{\citenamefont {Zhou}()}]{STCF}%
  \BibitemOpen
  \bibfield  {author} {\bibinfo {author} {\bibfnamefont {X.}~\bibnamefont
  {Zhou}},\ }in\ \href@noop {} {\emph {\bibinfo {booktitle} {{Workshop of the
  progress on the Super Tau Charm Factory, August 2020}}}}\BibitemShut
  {NoStop}%
\bibitem [{\citenamefont {Cortina~Gil}\ \emph {et~al.}(2021)\citenamefont
  {Cortina~Gil} \emph {et~al.}}]{NA62:2020pwi}%
  \BibitemOpen
  \bibfield  {author} {\bibinfo {author} {\bibfnamefont {E.}~\bibnamefont
  {Cortina~Gil}} \emph {et~al.} (\bibinfo {collaboration} {NA62}),\ }\href
  {\doibase 10.1007/JHEP02(2021)201} {\bibfield  {journal} {\bibinfo  {journal}
  {JHEP}\ }\textbf {\bibinfo {volume} {02}},\ \bibinfo {pages} {201} (\bibinfo
  {year} {2021})},\ \Eprint {http://arxiv.org/abs/2010.07644} {arXiv:2010.07644
  [hep-ex]} \BibitemShut {NoStop}%
\bibitem [{\citenamefont {Cata}\ and\ \citenamefont
  {Mateu}(2007)}]{Cata:2007ns}%
  \BibitemOpen
  \bibfield  {author} {\bibinfo {author} {\bibfnamefont {O.}~\bibnamefont
  {Cata}}\ and\ \bibinfo {author} {\bibfnamefont {V.}~\bibnamefont {Mateu}},\
  }\href {\doibase 10.1088/1126-6708/2007/09/078} {\bibfield  {journal}
  {\bibinfo  {journal} {JHEP}\ }\textbf {\bibinfo {volume} {09}},\ \bibinfo
  {pages} {078} (\bibinfo {year} {2007})},\ \Eprint
  {http://arxiv.org/abs/0705.2948} {arXiv:0705.2948 [hep-ph]} \BibitemShut
  {NoStop}%
\end{thebibliography}%

\end{document}